\documentclass[11pt,a4paper]{article}
\usepackage[T1]{fontenc}
\usepackage[utf8]{inputenc}
\usepackage[margin=2.5cm]{geometry}
\usepackage{amsmath,amssymb,amsthm}
\usepackage{mathtools}
\usepackage{algorithm}
\usepackage{algpseudocode}
\usepackage{booktabs}
\usepackage{graphicx}
\usepackage{hyperref}
\usepackage{cleveref}
\usepackage{tikz}
\usetikzlibrary{positioning, arrows.meta}
\newcommand{\alw}{\Box} 
\newcommand{\som}{\Diamond} 
\newcommand{\nex}{\raisebox{0.2em}{\scriptsize $\bigcirc$}}

\newcommand{\unt}{\,{\cal U}} 

\newtheorem{definition}{Definition}

\graphicspath{{./}{./fig/}}

\title{A Hierarchy of Composition Procedures for\\ Logic-Based Workflow Patterns}

\author{Radosław Klimek\\
\small AGH University of Krakow\\
Kraków, Poland\\
email: \url{rklimek@agh.edu.pl}}
\date{2026}

\begin{document}
\maketitle

\begin{abstract}
Logic-based workflow patterns capture recurring behavioural structures
together with reusable temporal specifications. Their composition raises
distinct problems as workflow structures become richer, including
instantiation, nesting, conditional execution, repeated occurrences, and
traceability of generated formulae. This paper proposes a hierarchy of
Pattern Composition ($\Pi C$) procedures, from $\Pi C_0$ to $\Pi C_4$,
that organizes these requirements as progressively richer compositional
capabilities. The hierarchy separates compositional complexity from the
expressive power of the temporal-logic fragment, represented by $TL_j$,
yielding a framework $\Pi C_i[TL_j]$ for developing logical workflow
specifications.
We develop activation-aware Pattern Composition ($\Pi C_2$), which
distinguishes the structural presence of a nested pattern from its
execution in a particular workflow context. We formalise its recursive
procedure, analyse its properties, and define an evaluation protocol for
conditional activation, conservativity under mandatory composition, and
specification growth. The hierarchy also characterizes selective
occurrence qualification for iterative patterns ($\Pi C_3$) and provenance
for tracing generated formulae to their sources ($\Pi C_4$). Finally, we
outline $\Pi C_{\mathrm{Assist}}$, an orthogonal mode in which AI proposes
and refines pattern expressions subject to the hierarchy's compositional
checks.
\end{abstract}

\section{Introduction}

Formal behavioural specification is a fundamental part of the modelling and
analysis of software and workflow-based systems~\cite{Broy-2013,Broy-2018}. 
Although concrete systems may differ considerably, many behavioural structures recur
across applications. Sequences of activities, alternatives, parallel
executions, synchronization, and other control-flow structures represent
reusable forms of behaviour rather than properties specific to a single
system.

Logic-based workflow patterns provide a reusable way of associating
recurring behavioural structures with their temporal specifications.
A pattern captures a characteristic form of behaviour together with
a set of logical formulae describing its intended properties. The key
principle is to formalise such behaviour once and reuse its specification
whenever the corresponding behavioural structure occurs, rather than
derive the required temporal formulae from scratch for each workflow.
Thus, the effort of defining and validating the logical properties is
incurred once, while the resulting specification can be instantiated
repeatedly with different actual activities and reused across different
workflows and applications. Logical patterns therefore provide both
reusable behavioural abstractions and reusable units of formal
specification.

The usefulness of individual patterns, however, depends fundamentally on
their composition. Compositionality is a well-established principle for
constructing and reasoning about complex systems from simpler
components~\cite{Tripakis-2016,Tripakis-2016b}. Realistic behaviours are
rarely represented by a single isolated pattern; instead, they are built
from multiple patterns combined into larger structures. \emph{Pattern
Composition}, denoted by $\Pi C$, applies this principle to logic-based
workflow patterns, providing a systematic mechanism for deriving complex
behavioural specifications from local and reusable pattern
specifications. Thus, composition enables the reuse of individual
logical patterns to scale towards the specification of increasingly
complex behaviours.

Not all forms of composition, however, pose the same problem.
Instantiating a flat pattern, embedding one pattern within another,
selecting conditionally executed branches, and reasoning about repeated
executions require different kinds of compositional information.
Increasing the structural complexity of a workflow therefore does not
merely produce a larger instance of the same composition problem, but
introduces qualitatively different requirements on the composition
mechanism. Nesting requires an interface between the internal behaviour
of a sub-pattern and its enclosing pattern; conditional execution
requires distinguishing structural presence from actual execution;
iteration introduces repeated occurrences and may require them to be
distinguished; and combinations of these mechanisms may introduce
further interactions between their respective composition requirements.

These observations motivate the central thesis of this work: Pattern
Composition should be organised as a hierarchy of procedures rather than
treated as a single universal composition mechanism. We introduce the
levels $\Pi C_0,\Pi C_1, \ldots$, progressing from flat and hierarchical
composition through conditional activation and iteration to their
combined use. The hierarchy identifies distinct mechanisms for composing
logical patterns and specifies their respective applicability conditions.

There is also another source of complexity that should be distinguished
from pattern composition itself: the expressive power of the logical
language used to specify behaviour. Temporal logic is a well-established
formalism~\cite{Emerson-1990,Manna-Pnueli-1992} 
for describing behavioural properties evolving over time and
provides a natural foundation for logical workflow specifications.
Temporal specifications, however, may range from relatively simple uses
of temporal modalities to formulae involving increasingly complex
nesting and interaction of temporal operators. Increasing compositional
complexity and increasing temporal expressiveness are therefore two
different dimensions of the specification problem.

This distinction also matters for automated reasoning. Although theorem
provers exist for temporal logic, the automated-reasoning ecosystem for
first-order logic appears substantially broader and more mature.
Consequently, whenever a temporal specification admits a suitable
translation into FOL, such a translation may provide access to a wider
range of reasoning tools. The translatability of temporal fragments to
FOL is therefore an important practical consideration alongside their
expressive power.

The broader objective is to bring together several complementary aspects
of logic-based behavioural specification: stable and reusable behavioural
patterns captured by logical formulae, different levels of
compositional complexity represented by $\Pi C_i$, different levels of
temporal expressiveness represented by $TL_j$, and the practical
availability of automated reasoning support. These aspects should be
considered jointly: reusable logical patterns provide the specification
units, Pattern Composition determines how such units can be combined,
the temporal fragment determines the expressive power of their
properties, and the available reasoning ecosystem influences how these
properties can be verified in practice. In particular, suitable
translations from temporal fragments to FOL may provide access to its
broader theorem-proving ecosystem. 

The main contributions of this work are as follows:
\begin{enumerate}

\item We organize Pattern Composition as a $\Pi C_0$--$\Pi C_4$
hierarchy, progressing from flat instantiation and hierarchical
composition through activation-aware composition and occurrence
qualification to provenance preservation. The hierarchy identifies the
additional compositional information required at each level.

\item We separate compositional complexity, represented by $\Pi C_i$,
from temporal-logic expressiveness, represented by $TL_j$. This provides
a two-dimensional framework $\Pi C_i[TL_j]$ in which composition
mechanisms and temporal expressiveness can be studied independently.

\item We develop activation-aware Pattern Composition ($\Pi C_2$),
which distinguishes the structural presence of a nested pattern from
its execution in a particular context. We formalise the procedure
recursively using parent--argument activation relations, with exclusive
choice as the principal case of conditional execution.

\item We analyse the properties of $\Pi C_2$ and define an evaluation
protocol based on systematically constructed pattern-expression
families. The protocol examines conditional-activation correctness,
conservativity under mandatory composition, and specification growth.

\item We introduce a provenance model for tracing generated formulae to
their originating pattern instances, composition rules, and contexts,
and outline $\Pi C_{\mathrm{Assist}}$ as an orthogonal mode for
AI-assisted proposal, validation, and refinement of pattern
expressions.

\end{enumerate}

\section{Background and Foundations}

The approach developed in this work builds on four complementary
foundations. First, recurring behavioural structures can be captured as
reusable logical specifications. Second, compositionality provides a
principle for constructing increasingly complex behavioural
specifications from such patterns. Third, temporal logic offers different
levels of behavioural expressiveness, ranging from simple temporal
properties to more involved combinations of temporal operators. Finally,
practical verification depends on the available automated-reasoning
ecosystems, which makes translation between logics---in particular from
temporal to first-order logic where feasible---an important consideration.
The following subsections discuss these four dimensions in turn.

\subsection{Behavioural Patterns as Reusable Logical Specifications}

Patterns constitute a well-established means of capturing recurring
structures and solutions in computing. Their use is particularly
prominent in software design and architecture, where a pattern
encapsulates knowledge that can be defined once and subsequently reused
in different systems and contexts~\cite{Gamma-etal-1995}.
The same principle applies to system behaviour. Recurrent control-flow
structures, such as sequences, alternatives, parallel execution,
synchronisation, and iteration, occur across different workflow and
process models and can therefore be treated as reusable behavioural
patterns~\cite{Aalst-etal-2003,Russell-etal-2016}.

In this work, behavioural patterns are considered from a logical
perspective. A pattern is not merely a recurring structural
configuration, but is associated with a predefined set of temporal
formulae characterising its intended behaviour. It can thus be viewed
as a local logical specification schema: its structure represents a
recurring behavioural situation, while its formulae specify the
properties expected of every valid instance of that situation. Such a
specification should capture both liveness properties, describing
behaviour that is eventually required to occur, and safety properties,
excluding undesirable behaviour~\cite{Kindler-1994,Alpern-Schneider-1985,Manna-Pnueli-1992}.

\begin{definition}[Behavioural pattern]
A behavioural pattern is a tuple
\[
\mathcal{P}=\langle op,Args,Spec,ini,fin\rangle,
\]
where:
\begin{itemize}
    \item $op$ is the pattern name (identifier);
    \item $Args=\{a_1,\ldots,a_n\}$ is a finite set of formal arguments
          representing atomic activities;
    \item $Spec\subset TL$ is a finite set of temporal formulae
          specifying the intended behaviour of the pattern;
    \item $ini$ and $fin$ are propositional formulae over $Args$
          representing the initial and final expressions of the pattern,
          respectively.
\end{itemize}
The conditions under which such a pattern constitutes a well-composed
logical specification are introduced below.
\end{definition}
A pattern $\mathcal{P}$ can be written schematically as
\[
P(a_1,\ldots,a_n).
\]
It constitutes a local specification entity, while its use in a
particular behavioural model follows the standard mechanism of formal
and actual arguments. The formal arguments serve as placeholders for
activity roles and may occur in the temporal formulae defining the
pattern. When the pattern is instantiated, they are replaced by actual
arguments representing concrete activities. Consequently, the same
pattern schema and its logical specification can be reused in different
behavioural contexts without modifying its definition.

Associated with $P$ is its logical specification, denoted by $P.pat()$,
consisting of temporal formulae that characterise the complete intended
behaviour of the pattern. The pattern additionally exposes distinguished
expressions describing its initial and final behaviour, $P.ini$ and
$P.fin$. A pattern can consequently be viewed at two related levels.
Its internal behavioural specification, represented by $P.pat()$,
describes the properties of the pattern itself, whereas $P.ini$ and
$P.fin$ constitute its externally visible compositional interface.
This distinction allows a pattern to remain a local specification entity
while providing the abstraction required for its subsequent use as a
component of a larger behavioural structure.

The logical specification associated with an individual behavioural
pattern should be internally consistent and should properly relate the
initial and final behaviour exposed by its interface. In particular,
the initial and final expressions should not hold simultaneously,
which gives the basic safety requirement
\[
\alw\neg(P.ini \land P.fin),
\]
while every initiation of the pattern should eventually lead to its
completion, expressed by the liveness requirement
\[
\alw(P.ini \rightarrow \som P.fin).
\]
Moreover, both interface expressions should represent reachable
behaviour rather than being trivially false. These requirements lead
to the following local notion of well-composedness.

\begin{definition}[Well-composed behavioural pattern]
A behavioural pattern
\[
\mathcal{P}=\langle op,Args,Spec,ini,fin\rangle
\]
is \emph{well-composed} if:
\begin{enumerate}
    \item
    $
    Spec \cup
    \left\{
        \alw\neg(ini \land fin),
        \alw(ini \rightarrow \som fin)
    \right\}
    $
    is satisfiable;
    
    \item $ini$ and $fin$ are reachable under the specification,
    i.e., there exist behaviours satisfying the specification in which
    the corresponding initial and final conditions occur.
\end{enumerate}
\end{definition}

Well-composedness is therefore a local property of an individual
behavioural pattern. It concerns the consistency of its internal logical
specification and the meaningfulness of the behavioural interface
through which the pattern can subsequently participate in larger
compositions. It does not, by itself, impose conditions on interactions
between different pattern instances; such conditions arise only when
patterns are composed into larger behavioural structures.

This locality is essential for composition: it allows the internal
properties of a pattern to be established independently of a particular
composition context, while composition can focus on the relations
between the participating pattern instances and their interfaces.

For example, consider a simple sequence pattern $Seq(a,b)$. Its
interface is given by
\[
Seq.ini \equiv a,
\qquad
Seq.fin \equiv b,
\]
while its logical specification invariably includes the liveness property
\[
\alw(a \rightarrow \som b)
\]

A central practical motivation for logical patterns is reuse. The
logical properties of a recurring behavioural structure are defined and
validated once and can subsequently be instantiated whenever that
structure occurs, rather than being reconstructed for every workflow.
Moreover, the definition and validation of such patterns may be carried
out by a logician or a specialist experienced in formal methods, while
their subsequent use can be delegated to analysts who need not derive
the underlying temporal formulae themselves. The resulting pattern
library can therefore separate the expertise required to construct
formal specifications from that required for their routine application,
supporting broader and repeated use of formal behavioural specification.

In summary, a well-composed behavioural pattern provides a reusable, 
context-independent logical specification that can be instantiated and 
composed into larger structures while preserving its local correctness.

\subsection{Compositionality in Behavioural Specification}

Compositionality is a fundamental principle for managing complexity in
computer science. Complex systems are routinely constructed from smaller
units that can be developed, understood, analysed, and reused as
identifiable entities. This principle underlies modular software design,
component-based systems, hierarchical modelling, and compositional
verification~\cite{Parnas-1972,Blume-Appel-1999,Tripakis-2016}.
More generally, it reflects a natural strategy for dealing with
complexity: complex entities are constructed from smaller entities with
well-defined roles and boundaries rather than treated monolithically.

Importantly, compositionality is not a single uniform operation.
Compositional verification, for example, relies on mechanisms such as
abstraction, simplification, hiding, and property-preserving
equivalences, depending on what is composed and which properties are to
be preserved~\cite{Henzinger-etal-1998,Malik-etal-2023}. This illustrates
a more general principle: extending compositional capability may require
new mechanisms rather than repeated application of the same composition
operation.

For behavioural specification, this principle naturally complements the
reuse provided by logical patterns. Individual behavioural patterns are
local, reusable specification units; composition makes it possible to
construct larger behavioural specifications from such units. Instead of
deriving a monolithic temporal specification for every workflow, we aim
to obtain it systematically from logical specifications already defined
and validated for its constituent behavioural patterns.

This leads to \emph{Pattern Composition}, denoted by $\Pi C$, understood
as a systematic procedure for constructing the logical specification of
a composite behaviour from instantiated specifications of its
constituent behavioural patterns. The logical definition of an
individual pattern remains local and reusable, while the composition
procedure determines how its instances contribute to the specification
of the whole.

A \emph{pattern expression} is an instantiated behavioural pattern whose
actual arguments are either atomic activities or, whenever admissible,
other pattern expressions. Thus, pattern expressions provide the
structural representation of behaviours to which Pattern Composition is
applied. For example,
\[
P(a,b,c)
\]
is a flat pattern expression, whereas
\[
P(a,Q(b,c),d)
\]
is a hierarchical pattern expression containing the nested expression
$Q(b,c)$.

Abstractly, Pattern Composition can be represented as a mapping
\[
\Pi C : \mathcal{E} \longrightarrow 2^{TL},
\]
where $\mathcal{E}$ denotes the class of admissible behavioural pattern
expressions. For an expression $E\in\mathcal{E}$,
\[
\Pi C(E)=Spec_E,
\]
where $Spec_E\subseteq TL$ is the temporal specification generated for
$E$.

Different compositional capabilities can then be associated with
increasingly rich classes of admissible pattern expressions,
\[
\mathcal{E}_0 \subseteq
\mathcal{E}_1 \subseteq
\mathcal{E}_2 \subseteq \cdots .
\]
Intuitively, each extension of the admissible expression class may
introduce structures that cannot be handled by the preceding composition
mechanism and therefore require additional compositional information.
The corresponding composition procedures will be formalised later as
the hierarchy of $\Pi C_i$ levels.

In the simplest case, all actual arguments are atomic activities.
Composition then essentially reduces to instantiation: formal arguments
are replaced by the corresponding actual activities and the predefined
logical specification is instantiated accordingly. No additional
compositional mechanism is required.

This is no longer sufficient when an actual argument represents a
structured behaviour, as in the hierarchical expression above. The
nested expression $Q(b,c)$ has its own internal structure and logical
specification and cannot be treated simply as an atomic activity.
Hierarchical composition therefore requires an additional mechanism
that provides an appropriate abstraction of the nested behaviour at its
composition boundary. Moreover, not every nesting is necessarily
admissible; composition must respect the behavioural roles and
interfaces of the participating patterns.

Further behavioural constructs introduce different requirements.
Conditional behaviour raises the distinction between structural
presence and actual execution: a component may occur in the behavioural
structure without being active in a particular execution. Iteration
introduces repeated executions of the same behavioural component and
therefore raises the problem of distinguishing its occurrences.

The progression can be summarised as follows:
\[
\begin{array}{rcl}
\text{atomic arguments}
    &\Longrightarrow& \text{instantiation},\\
\text{nested behaviour}
    &\Longrightarrow& \text{structural abstraction},\\
\text{conditional behaviour}
    &\Longrightarrow& \text{activation information},\\
\text{iterated behaviour}
    &\Longrightarrow& \text{occurrence identity}.
\end{array}
\]

This progression captures the key point: increasing compositional
complexity does not merely produce larger instances of the same problem.
Each step introduces a new kind of information that the preceding
mechanism does not provide. A new composition procedure is therefore
justified when a class of behavioural structures introduces such a new
requirement. The hierarchy developed in the following section follows
precisely this principle: successive levels extend compositional
capability by addressing qualitatively new requirements rather than
merely increasing syntactic size or nesting depth.

\subsection{Temporal Expressiveness in Behavioural Specification}

The complexity of Pattern Composition is not the only dimension relevant
to behavioural specification. A separate dimension concerns the temporal
language in which behavioural properties are expressed. The two
dimensions should not be conflated: increasing the structural capability
of composition does not necessarily require a corresponding increase in
the complexity or expressive power of the temporal language.

We restrict attention to propositional linear-time temporal logic over
discrete executions. To keep the temporal dimension focused on the
requirements relevant to behavioural patterns, we consider future-time
operators only. Past-time operators and branching-time temporal logics,
although providing further dimensions of temporal expressiveness, are
outside the scope of this work.

Temporal operators provide different ways of
relating the occurrence of activities along an execution. In the setting
considered here, the basic temporal vocabulary relies primarily on the
``eventually'' and ``always'' operators, denoted by $\som$ and $\alw$,
respectively. They naturally express properties such as eventual
occurrence,
\[
\som p,
\]
response,
\[
\alw(p \rightarrow \som q),
\]
and invariance or safety,
\[
\alw\neg(p \land q).
\]
These forms already cover many of the liveness and safety requirements
associated with the behavioural patterns considered in this work.

Even a vocabulary based only on $\som$ and $\alw$ admits temporal
formulae of different structural complexity. Simple occurrences of a
temporal operator can be combined into response properties, while
nesting permits properties such as
\[
\alw\som p
\qquad\text{and}\qquad
\som\alw p,
\]
expressing recurrence and persistence, respectively. More involved
formulae may combine several such structures, for example
\[
\alw(p \rightarrow \som(q \land \alw r)).
\]
Thus, temporal specification complexity may increase through the
operators being used, their combinations, and their nesting depth.

These dimensions should not, however, be identified automatically with
expressive power. Temporal nesting, syntactic complexity, and expressive
power are related but distinct notions. In particular, adding another
level of nesting does not in general justify claiming a new level of
expressiveness without a corresponding formal result. This distinction
is important here because the increasingly advanced forms of Pattern
Composition developed in this work do not currently require increasingly
deep nesting of temporal operators.

Other temporal operators introduce qualitatively different specification
capabilities. The next operator $\nex$ refers to the immediate successor
of the current state:
\[
\nex p.
\]
Unlike eventuality, which requires $p$ to hold at some future point,
$\nex p$ requires it at the next point of the execution. This makes
$\nex$ particularly dependent on the discrete step structure and
distinguishes it from the stutter-invariant fragment of LTL
without next~\cite{Peled-Wilke-1997}. The operator is therefore best
viewed not simply as a ``more complex'' version of $\som$ or $\alw$,
but as providing a different form of temporal resolution.

The until operator $\unt$ introduces another, stronger form of temporal
structure. A formula
\[
p \unt q
\]
requires not only that $q$ eventually occur, but also constrains the
behaviour preceding that occurrence: $p$ must hold until the relevant
occurrence of $q$. In standard LTL, eventuality can itself be defined
using until,
\[
\som p \equiv \top \unt p,
\]
and always can subsequently be obtained by duality,
\[
\alw p \equiv \neg\som\neg p.
\]
Thus, the $\som/\alw$ vocabulary used by our current behavioural
patterns represents a restricted temporal fragment rather than the full
range of standard LTL constructs.

The distinction becomes particularly significant when until operators
are nested. Unlike an informal progression based merely on formula size,
the nesting depth of $\unt$ has a precise theoretical foundation:
successive until depths form a strict hierarchy of expressive power in
linear temporal logic~\cite{Etessami-Wilke-1996}. Related results also
show that nesting of next and until gives rise to meaningful temporal
hierarchies~\cite{Kucera-Strejcek-2002}. Consequently, richer temporal
structure is not merely a matter of writing longer formulae; for
appropriate operator fragments it may correspond to genuine differences
in what temporal properties can be expressed.

For the purposes of this work, these observations suggest a progression
of temporal specification capabilities rather than a universal ordering
of temporal operators:
\[
\begin{array}{rcl}
\text{basic }\som/\alw\text{ properties}
    &\Longrightarrow& \text{occurrence and invariance},\\
\text{combined }\som/\alw\text{ properties}
    &\Longrightarrow& \text{response and related dependencies},\\
\text{nested }\som/\alw\text{ structures}
    &\Longrightarrow& \text{recurrence, persistence, and richer dependencies},\\
\nex
    &\Longrightarrow& \text{immediate-successor sensitivity},\\
\unt
    &\Longrightarrow& \text{constraints over temporal intervals},\\
\text{nested }\nex/\unt\text{ structures}
    &\Longrightarrow& \text{further temporal distinctions}.
\end{array}
\]

This progression is deliberately not presented as a strict hierarchy of
expressive power. Its purpose is to distinguish increasingly involved or
qualitatively different temporal specification requirements. Some of
these distinctions, notably those based on until depth, correspond to
formally established expressiveness hierarchies, whereas others identify
different temporal capabilities or syntactic structures.

Let $TL_0$ denote the restricted temporal fragment sufficient for the
behavioural pattern specifications considered in this work. Possible
extensions of the temporal language can be represented abstractly as
\[
TL_0 \subseteq TL_1 \subseteq TL_2 \subseteq \cdots ,
\]
where richer fragments may admit additional temporal constructs or
more involved combinations of operators. We do not fix these subsequent
levels here: their definition should be driven by temporal properties
actually required by future behavioural specifications rather than by
an artificial increase in operator nesting.

This temporal dimension is independent of the hierarchy of Pattern
Composition. Hierarchical composition may require structural abstraction,
conditional composition may require activation information, and
iteration may require occurrence information while the generated
properties remain within the same temporal fragment. Conversely, the
temporal language may be extended without changing the composition
mechanism. In the present work, we therefore advance primarily along the
compositional dimension while retaining the temporal specification within
$TL_0$; richer temporal fragments remain an independent direction of
extension. This independence is a key methodological principle: it allows
us to study compositional mechanisms independently of the expressive
richness of the temporal fragment, and to extend the temporal language
without necessarily revising the composition hierarchy.

\subsection{Automated Reasoning Ecosystems and Logic Translation}

Formal behavioural specifications are intended not only to describe
system behaviour, but also to support automated deduction. This
introduces a practical dimension that complements both compositional
complexity and temporal expressiveness: different logical formalisms are
supported by different reasoning ecosystems, with different proof
procedures, input languages, tool infrastructures, and levels of
automation and maturity. A logic that provides a natural and compact
language for expressing behavioural properties need not therefore offer
the most suitable reasoning infrastructure for every deductive task.
The choice of a specification language should not unnecessarily restrict
the reasoning technology that can subsequently be employed.

Temporal logic is the natural specification formalism in the present
setting because the properties of interest describe behaviour evolving
over time. Dedicated reasoning procedures and tools for temporal logics
are well established. For propositional linear temporal logic, for
example, systems such as TRP++ provide temporal-resolution-based
reasoning, while TeMP and TSPASS illustrate corresponding developments
for fragments of first-order temporal logic. The temporal reasoning
landscape therefore provides specialised deductive machinery, although
its tools are distributed across particular temporal formalisms and
reasoning procedures.

We therefore treat temporal expressiveness as an independent dimension,
since it determines not only what behavioural properties can be
specified, but also which automated reasoning routes are available.

Our interest is primarily deductive: generated logical specifications
are intended to serve as objects of automated inference rather than only
as properties to be checked against a given system model. From this
perspective, first-order logic is particularly attractive because it is
supported by a broad and mature ecosystem of automated theorem provers.
Systems such as Vampire, E, iProver, and Prover9 coexist with standardised
problem representations, extensive benchmark libraries, and common
infrastructures provided by TPTP~\cite{Sutcliffe-2017}. Importantly, this
ecosystem has also supported long-established competitive evaluation
through the CADE ATP System Competition (CASC)~\cite{Sutcliffe-2021}.
This combination of mature provers and shared infrastructure provides a
strong practical motivation for exploiting first-order reasoning whenever
a suitable translation of the temporal specification is available.

This motivates a deliberately flexible reasoning strategy. Behavioural
properties should be specified in a logic that represents them naturally,
while deduction should exploit a reasoning ecosystem appropriate to the
resulting problem. Reasoning may therefore be performed directly in the
source temporal logic when appropriate deductive machinery is available.
Alternatively, when a controlled translation to another logic is feasible
and advantageous, the translated specification may be submitted to the
corresponding reasoning tools. First-order logic constitutes a particularly
relevant target in this respect. The resulting reasoning strategy can
therefore follow two principal routes:
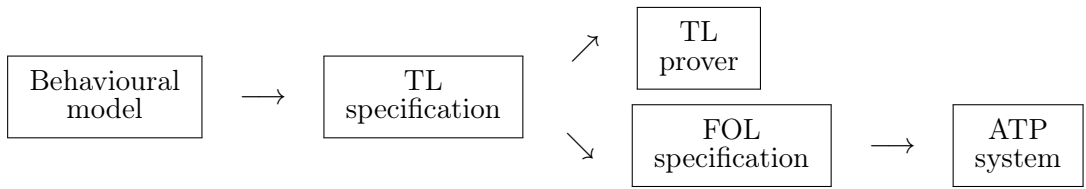
\[
\boxed{\begin{array}{c}
\text{Behavioural}\\[-1mm]
\text{model}
\end{array}}
\quad\longrightarrow\quad
\boxed{\begin{array}{c}
\text{TL}\\[-1mm]
\text{specification}
\end{array}}
\quad
\begin{array}{c}
\hspace*{-42mm}
\nearrow\quad
\boxed{\begin{array}{c}
\text{TL}\\[-1mm]
\text{prover}
\end{array}}
\\[5mm]
\searrow\quad
\boxed{\begin{array}{c}
\text{FOL}\\[-1mm]
\text{specification}
\end{array}}
\quad\longrightarrow\quad
\boxed{\begin{array}{c}
\text{ATP}\\[-1mm]
\text{system}
\end{array}}
\end{array}
\]
The first route preserves the source temporal formalism and employs
deductive machinery available for that logic. The second introduces a
controlled translation to a target logic, with FOL being particularly
relevant because of its mature ATP infrastructure. Translation is thus
not intended to replace temporal logic as the specification language,
but to provide an additional deductive route whenever such a route is
feasible and advantageous.

Temporal-to-first-order translation has a well-established logical
basis. Let temporal formulae be interpreted over a linear order of time
points $(T,<)$, and let $ST_t(\varphi)$ denote the \emph{standard
translation} of a temporal formula $\varphi$ evaluated at time point
$t$. For the basic future operators used extensively in our behavioural
specifications, the translation is comparatively direct:
\[
ST_t(\som\varphi)
=
\exists u\,
\bigl(t \leq u \land ST_u(\varphi)\bigr),
\]
and
\[
ST_t(\alw\varphi)
=
\forall u\,
\bigl(t \leq u \rightarrow ST_u(\varphi)\bigr).
\]
Thus, the familiar intuition
\[
\som \sim \exists,
\qquad
\alw \sim \forall
\]
is useful only as shorthand. Temporal operators are not replaced by
quantifiers alone; their temporal meaning is recovered by quantifying
over explicitly ordered time points.

For example, the response property
\[
\alw(p\rightarrow\som q)
\]
has the first-order representation
\[
\forall t\,
\left(
P(t)\rightarrow
\exists u\,
(t\leq u\land Q(u))
\right).
\]
What is implicit in the temporal formulation---quantification over the
future of the current point---becomes explicit in the first-order
representation. Even this common behavioural property already introduces
a quantifier dependency of the form $\forall\exists$.

Other temporal operators illustrate why translation should not be
understood as a mechanical replacement of modalities by quantifiers.
For discrete executions, the next operator depends on the immediate
successor structure. Schematically,
\[
ST_t(\nex\varphi)
=
\exists u\,
\bigl(Next(t,u)\land ST_u(\varphi)\bigr),
\]
where $Next(t,u)$ identifies $u$ as the immediate successor of $t$.
The until operator introduces a different kind of structure:
\[
ST_t(\varphi\unt\psi)
=
\exists u
\left(
t\leq u
\land ST_u(\psi)
\land
\forall v
\bigl(
t\leq v<u
\rightarrow ST_v(\varphi)
\bigr)
\right).
\]
Here, translation must identify a point at which $\psi$ holds while
simultaneously constraining the interval preceding that point. The
resulting first-order representation is therefore structurally more
involved than that of a basic eventuality or invariance property.

Increasing temporal complexity does not, within the propositional
linear-time setting considered here, imply a loss of translatability
to first-order logic. It may, however, substantially affect the
structure of the resulting first-order representation. Simple uses of
$\som$ and $\alw$ give rise to comparatively direct quantification over
ordered time points; combinations and nesting of temporal operators may
introduce more involved quantifier dependencies; $\nex$ depends on the
successor structure of discrete time; and $\unt$ requires constraints
over an interval of temporal positions. These cases should therefore
not be interpreted as a simple hierarchy of increasingly
``difficult'' temporal operators. Rather, different temporal constructs
introduce different representational requirements, which may in turn
affect the subsequent deductive task.

This leads to an important distinction. For practical automated
deduction, at least three questions have to be separated:
\[
\begin{array}{rcl}
\text{translatability}
&:&
\text{can the temporal property be represented in the target logic?}\\
\text{translation complexity}
&:&
\text{what additional logical structure does the translation introduce?}\\
\text{deductive tractability}
&:&
\text{how effectively can available provers reason over the result?}
\end{array}
\]
These questions are related but should not be conflated. A translation
may be theoretically available while producing a representation that is
unattractive for practical automated deduction. Conversely, restricted
temporal fragments may admit comparatively transparent translations
whose resulting formulae are well suited to existing automated
reasoning technology. Translation should consequently be treated as a
reasoning opportunity rather than as an obligatory processing step.

The relationship between temporal and first-order logic is not
incidental. Classical results, most notably Kamp's
theorem~\cite{Kamp-1968}, establish an expressive correspondence between
temporal logic with Until and Since over linear orders and an
appropriate monadic fragment of first-order logic with order. More
generally, van Benthem's model-theoretic account of modal and temporal
logics shows how such languages can be systematically related to
first-order logic through relational semantics and standard
translation~\cite{vanBenthem-1991-LogicOfTime}.

In this view, temporal logic provides a natural and compact language for
describing temporal behaviour, whereas first-order logic makes time
points and their ordering explicit. The two formalisms should therefore
not be placed in a simplistic hierarchy of logical ``strength''.
Rather, they provide different representations of temporal structures,
which can be exploited when selecting an appropriate route for
automated deduction.

For the framework developed in this work, the resulting methodological
principle is pragmatic. We retain temporal logic as the primary language
of behavioural specification, but do not bind automated deduction to a
single logic or prover family. Direct temporal reasoning remains
possible where suitable deductive machinery is available; controlled
translation provides an additional route where another reasoning
ecosystem is advantageous. In particular, translation to first-order
logic provides access to a broad ATP ecosystem whenever the temporal
fragment and the resulting translation make this route practically
attractive. The choice of temporal fragment is therefore relevant not
only to what behavioural properties can be expressed, but also to which
automated-deduction routes remain effectively available.

\section{A Hierarchy of Pattern Composition Procedures}

The conceptual dimensions introduced in the preceding sections provide
the basis for a hierarchy of Pattern Composition procedures. Rather than
treating nesting, conditional activation, and iteration as variants of a
single composition problem, we organise them into successive levels of
composition capability. Each level addresses a qualitatively new
compositional requirement while retaining the mechanisms introduced at
the preceding levels.

The resulting hierarchy progresses from direct pattern instantiation
through structural abstraction, activation-sensitive composition, and
occurrence-sensitive reasoning, ultimately combining these mechanisms
within a single composition procedure:
\[
\boxed{\text{Instantiation}}
\;\longrightarrow\;
\boxed{\text{Structural abstraction}}
\;\longrightarrow\;
\boxed{\text{Activation context}}
\;\longrightarrow\;
\boxed{\text{Occurrence identity}}
\;\longrightarrow\;\cdots
\]

The central idea of the proposed hierarchy is that increasing behavioural
complexity introduces qualitatively different composition problems.
Successive procedures therefore extend the applicability domain of
Pattern Composition step by step:
\[
\Pi C_0
\xrightarrow{\text{nesting}}
\Pi C_1
\xrightarrow{\text{conditional activation}}
\Pi C_2
\xrightarrow{\text{iteration}}
\Pi C_3
\xrightarrow{\text{combined composition}}
\Pi C_4.
\]

The hierarchy should not be interpreted merely as a sequence of
increasingly elaborate algorithms. Each transition introduces a
qualitatively different composition problem and removes a corresponding
restriction of the preceding level. The final level addresses the
interaction of mechanisms that can first be considered separately at
the preceding levels. An informal characterisation of these problems is:
\[
\begin{array}{lll}
\Pi C_0 &: & \textit{How is a flat pattern instantiated?}\\
\Pi C_1 &: & \textit{How is a nested pattern represented?}\\
\Pi C_2 &: & \textit{When is a nested pattern activated?}\\
\Pi C_3 &: & \textit{Which occurrence of a repeated pattern is considered?}\\
\Pi C_4 &: & \textit{How do these mechanisms interact?}
\end{array}
\]

The individual levels can be characterised more formally by
corresponding classes of admissible pattern expressions
\[
\mathcal{E}_0 \subseteq
\mathcal{E}_1 \subseteq
\mathcal{E}_2 \subseteq \cdots
\]
and composition procedures
\[
\Pi C_i : \mathcal{E}_i \longrightarrow 2^{TL}.
\]
Each extension of $\mathcal{E}_i$ admits a class of behavioural
structures requiring compositional information not provided at the
preceding level. The hierarchy of procedures therefore corresponds to
a progressive extension of the behavioural structures to which Pattern
Composition can be applied.

The hierarchy is intentionally open-ended. The levels $\Pi C_0$ through
$\Pi C_4$ capture the compositional requirements identified in the
present work; further levels may be introduced when additional classes
of behavioural structures require qualitatively new compositional
information.

The following subsections define each level in turn, starting with the
simplest case and progressively removing restrictions to cover richer
behavioural structures.

\begin{figure}[t]
\centering
\begin{tikzpicture}[
  level/.style={
    draw, rounded corners, thick, fill=blue!6,
    text width=2.1cm, minimum width=2.45cm,
    minimum height=1.05cm, inner sep=3pt,
    align=center, font=\small
  },
  flow/.style={-{Latex[length=2mm]}, thick},
  link/.style={densely dashed, gray}
]

\node[level] (pc0) at (1.225,0)
  {$\Pi C_0$\\Instantiation};
\node[level] (pc1) at (4.225,0)
  {$\Pi C_1$\\Structural abstraction};
\node[level] (pc2) at (7.225,0)
  {$\Pi C_2$\\Activation context};
\node[level] (pc3) at (10.225,0)
  {$\Pi C_3$\\Occurrence identity};
\node[level] (pc4) at (13.225,0)
  {$\Pi C_4$\\Provenance};

\draw[flow] (pc0.east) -- (pc1.west);
\draw[flow] (pc1.east) -- (pc2.west);
\draw[flow] (pc2.east) -- (pc3.west);
\draw[flow] (pc3.east) -- (pc4.west);

\draw[flow] (0.1,0.95) --
  node[above, font=\small] {increasing compositional information}
  (14.35,0.95);

\node[
  draw, rounded corners, thick, fill=orange!10,
  text width=14.1cm, minimum width=14.7cm,
  minimum height=0.85cm, inner sep=3pt,
  align=center, font=\small
] (assist) at (7.35,-1.45) {
  $\Pi C_{\mathrm{Assist}}$:
  proposes and refines expressions; the applicable hierarchy level
  checks them and returns feedback
};

\foreach \n in {pc0,pc1,pc2,pc3,pc4}
  \draw[link] (\n.south) -- (\n.south |- assist.north);

\draw[flow] (0.1,-2.8) --
  node[above, font=\small] {increasing temporal-logic expressiveness}
  (11.9,-2.8);

\node[font=\scriptsize, below] at (1.2,-2.8)
  {less expressive fragments};
\node[font=\scriptsize, below] at (10.6,-2.8)
  {more expressive fragments};

\node[
  draw, rounded corners, thick, fill=green!10,
  text width=2.1cm, minimum width=2.45cm,
  minimum height=0.75cm, inner sep=3pt,
  align=center, font=\small
] (joint) at (13.225,-2.8) {$\Pi C_i[TL_j]$};

\draw[flow] (11.9,-2.8) -- (joint.west);

\end{tikzpicture}
\caption{Pattern Composition hierarchy and its independent dimensions.
Each successive level adds compositional information. The orthogonal
AI-assisted mode can propose and refine expressions checked by the
applicable level. The temporal-logic fragment is selected independently,
yielding configurations $\Pi C_i[TL_j]$.}
\label{fig:pc-hierarchy}
\end{figure}
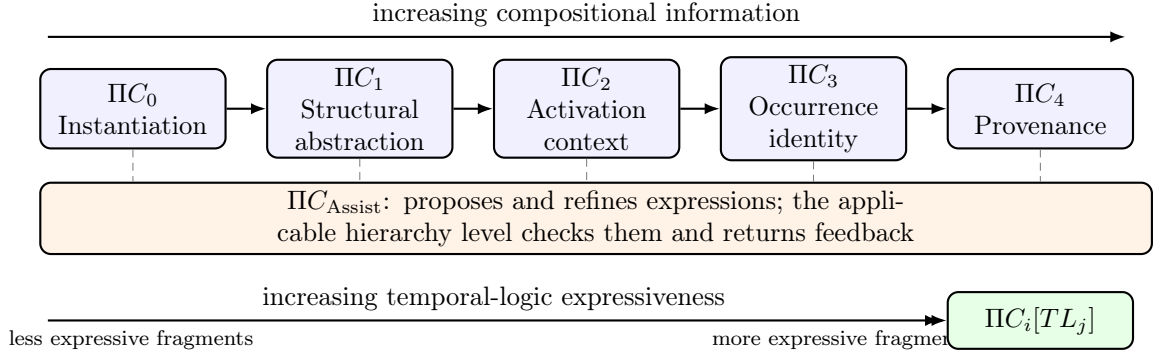

Figure~\ref{fig:pc-hierarchy} summarizes the Pattern Composition
hierarchy, its interaction with the orthogonal AI-assisted mode, and
the independent temporal-logic dimension.

\subsection{$\Pi C_0$: Flat Pattern Composition}

The base level, $\Pi C_0$, concerns local behavioural structures that
can be represented by a single pattern whose actual arguments are atomic
activities. A typical pattern expression has the form
\[
P(a_1,\ldots,a_n),
\]
where each argument $a_i$ is an atomic activity rather than another
pattern expression.

At this level, Pattern Composition does not yet require any additional
mechanism for handling internal structure, execution context, or repeated
occurrences. Its essential task is to identify the behavioural pattern
that matches the analysed structure and then instantiate its predefined
logical specification. Thus, before logical reuse takes place, the
analysed behavioural fragment must be recognised as an instance of an
appropriate pattern and its activities must conform to the structural
roles defined by that pattern.

Once the base pattern has been identified, composition reduces to the
standard formal--actual argument mechanism. If
\[
P(x_1,\ldots,x_n)
\]
is the predefined pattern schema and
\[
P(a_1,\ldots,a_n)
\]
is its concrete instance, then the generated specification is obtained
by substituting the actual activities for the corresponding formal
arguments in $P.pat()$. Schematically,
\[
P.pat(x_1,\ldots,x_n)
\;\xrightarrow{
x_1/a_1,\ldots,x_n/a_n
}\;
P.pat(a_1,\ldots,a_n).
\]

The simplicity of $\Pi C_0$ should therefore not be confused with the
absence of analysis. Pattern identification and structural conformance
must precede instantiation. What is absent at this level is additional
\emph{compositional} machinery: there is no nested behaviour to abstract,
no conditional activation to represent, and no repeated occurrence to
distinguish.

This level is also meaningful from a modelling perspective. Behavioural
analysis frequently begins locally, with individual recurring structures
whose properties can be considered independently before interactions and
larger structural dependencies are introduced. $\Pi C_0$ captures this
local stage and provides the baseline against which subsequent levels of
Pattern Composition can be characterised.

Accordingly,
\[
\Pi C_0
\quad=\quad
\text{pattern identification}
\;+\;
\text{direct instantiation},
\]
while the transition to $\Pi C_1$ occurs precisely when an actual
argument ceases to be atomic and becomes a structured behavioural
component. At that point, direct substitution alone is no longer
sufficient.

\subsection{$\Pi C_1$: Hierarchical Pattern Composition}

The first genuine compositional extension arises when an actual argument
of a behavioural pattern is no longer restricted to an atomic activity,
but may itself be a pattern expression. For example,
\[
P(a,Q(b,c),d)
\]
contains the nested expression $Q(b,c)$ as an argument of $P$.
Consequently, the behavioural structure can no longer be handled by the
direct formal--actual substitution used at level $\Pi C_0$.

Pattern expressions provide a compact structural representation of such
hierarchical behaviours. A pattern expression can be viewed as a
\emph{tree-like expression}: a recursively constructed term whose leaves
are atomic activities and whose internal nodes are behavioural patterns.
Subject to the admissibility conditions of the pattern formalism, its
structure may be described schematically by
\[
E ::= a
\mid
P(E_1,\ldots,E_n),
\]
where $a$ is an atomic activity and $P$ is a behavioural pattern.
An expression such as
\[
P(a,Q(b,c),R(d,e))
\]
therefore provides a linear notation for an underlying tree-like
behavioural structure while preserving the boundaries of its constituent
patterns.

This additional structural capability creates a new compositional
problem. A nested pattern such as $Q$ has its own logical specification,
which must remain part of the specification of the complete behaviour.
At the same time, the enclosing pattern $P$ cannot treat the complete
internal specification of $Q$ as though it were an atomic actual
argument. Instead, it requires an appropriate abstraction of the nested
behaviour at the composition boundary.

This role is played by the \emph{consolidated expression}, denoted by
$ConsEx(Q)$. The consolidated expression represents the nested pattern
to its parent using the behavioural information required at the
composition boundary. Hierarchical composition therefore introduces an
important distinction between
\[
\boxed{\text{internal behavioural specification}}
\qquad\text{and}\qquad
\boxed{\text{compositional interface}}.
\]
The former preserves the logical properties of the nested pattern,
whereas the latter provides the representation through which that
pattern participates in the specification of its enclosing pattern.
Thus, $ConsEx$ is an abstraction mechanism for composition rather than
a replacement for the internal specification of the nested behaviour.

Schematically, processing a nested pattern produces
\[
Q(b,c)
\longrightarrow
\bigl(Q.pat(),\,ConsEx(Q)\bigr).
\]
The first component contributes the local logical properties of $Q$ to
the resulting specification, while the second supplies the
representation required for instantiating the parent pattern. For
\[
P(a,Q(b,c),d),
\]
the parent can consequently be processed using
\[
P\bigl(a,ConsEx(Q),d\bigr),
\]
while the specification generated for $Q$ is retained as part of the
specification of the complete expression.

This mechanism naturally leads to bottom-up composition. Nested pattern
expressions are processed before their enclosing patterns, so that their
consolidated expressions are available when the corresponding parent is
instantiated. At the algorithmic level, $\Pi C_1$ can therefore be
summarised by the following processing scheme:
\[
\boxed{\text{nested expression}}
\;\longrightarrow\;
\boxed{\text{local specification}}
\;\longrightarrow\;
\boxed{\text{structural abstraction}}
\;\longrightarrow\;
\boxed{\text{enclosing-pattern instantiation}}.
\]
The procedure is applied recursively from the innermost patterns towards
the root of the pattern expression. The detailed construction of
consolidated expressions and the complete hierarchical composition
algorithm are given in~\cite{Klimek-2019-LAMP}. Here, this procedure is recast
as level $\Pi C_1$ of the proposed hierarchy in order to expose the
specific compositional capability that it provides: structural
abstraction of nested behaviour.

Accordingly,
\[
\Pi C_1 :
\mathcal{E}_1
\longrightarrow
2^{TL},
\]
where $\mathcal{E}_1$ extends the flat expressions of $\mathcal{E}_0$
with admissible hierarchical pattern expressions. The transition from
$\Pi C_0$ to $\Pi C_1$ is therefore not merely an increase in expression
size or nesting depth. It removes the restriction that actual arguments
must be atomic and introduces the first additional compositional
mechanism required by the hierarchy: a structural abstraction through
which nested behaviour can be represented to its parent.

\subsection{$\Pi C_2$: Activation-Aware Pattern Composition}

Hierarchical representation does not resolve every composition problem.
Level $\Pi C_1$ provides a structural abstraction through which a nested
pattern can be represented to its enclosing pattern. It does not,
however, distinguish between the structural presence of that pattern and
its execution in a particular behaviour. This distinction becomes
essential whenever the enclosing control-flow structure permits only
conditional execution of its arguments.

The fundamental observation underlying $\Pi C_2$ is therefore
\[
\boxed{
\text{structural presence}
\not\equiv
\text{behavioural activation}
}.
\]
A nested pattern may belong to the pattern expression, contribute its
structural abstraction to the enclosing pattern, and yet remain
unexecuted in a particular run.

Exclusive choice provides the simplest example. Consider
\[
XOR(P,Q),
\]
where both $P$ and $Q$ are nested behavioural patterns. Structurally,
both alternatives belong to the expression and must therefore be
represented during composition. Semantically, however, an execution of
the exclusive choice selects exactly one of them.

This difference becomes problematic if the local specifications of the
nested patterns independently impose unconditional existence
requirements such as
\[
Exist(P.ini),
\qquad
Exist(Q.ini).
\]
Combined with the semantics of exclusive choice, the resulting global
specification may then contain, schematically,
\[
Exist(P.ini)
\land
Exist(Q.ini)
\land
ExactlyOne(P,Q),
\]
which may be inconsistent. Even when only one alternative is nested, as
in
\[
XOR(P,a),
\]
an unconditional requirement $Exist(P.ini)$ may semantically force the
nested branch and thereby eliminate the alternative represented by $a$.
Thus, the problem is not exclusive choice itself, but the unconditional
activation of a local pattern independently of the context in which it
occurs.

This exposes a compositional requirement that is qualitatively different
from the structural abstraction introduced by $\Pi C_1$. The two levels
answer different questions:
\[
\begin{array}{rcl}
\Pi C_1 &:&
\textit{How is a nested pattern represented to its enclosing pattern?}\\[1mm]
\Pi C_2 &:&
\textit{Under what conditions is that nested pattern required to execute?}
\end{array}
\]
Representation and activation must therefore be treated separately.

The corresponding design principle is
\[
\boxed{
\text{local semantics}
\quad\perp\quad
\text{activation semantics}
}.
\]
A nested pattern should define the behaviour that must hold
\emph{if the pattern is executed}, whereas the enclosing pattern should
determine whether its execution is required in the current behavioural
context. For a mandatory sequential component, execution may be required
unconditionally; for an exclusive alternative, execution is conditional;
parallel and iterative contexts may impose still different activation
requirements.

This principle does not necessarily require the introduction of an
explicit proposition such as $Act(P)$. One possible realisation is to
guard local requirements by an activation condition,
\[
Act(P) \rightarrow Exist(P.ini),
\]
with the enclosing pattern constraining the activation of its children.
For example,
\[
ExactlyOne(Act(P),Act(Q))
\]
could capture an exclusive choice. An alternative, and potentially less
intrusive, realisation is to retain the existing pattern interface and
make the propagation of existence requirements context-sensitive: a
nested pattern contributes its local behavioural specification, while
the enclosing pattern determines whether $Exist(P.ini)$ must be imposed.
The precise realisation is developed later; at this level, activation
awareness is the essential compositional capability.

Accordingly, $\Pi C_2$ extends $\Pi C_1$ conservatively. Consolidated
expressions and bottom-up processing remain responsible for structural
representation. The additional mechanism concerns the propagation of
execution requirements through the enclosing pattern structure. At the
algorithmic level, the extension can be summarised as
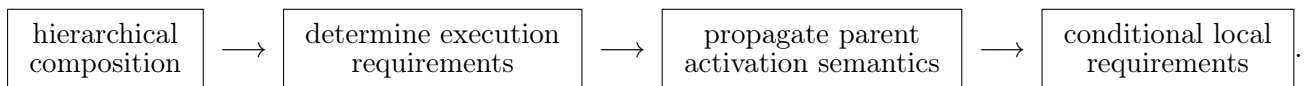
\[
\boxed{\begin{array}{c}
\text{hierarchical}\\[-1mm]
\text{composition}
\end{array}}
\;\longrightarrow\;
\boxed{\begin{array}{c}
\text{determine execution}\\[-1mm]
\text{requirements}
\end{array}}
\;\longrightarrow\;
\boxed{\begin{array}{c}
\text{propagate parent}\\[-1mm]
\text{activation semantics}
\end{array}}
\;\longrightarrow\;
\boxed{\begin{array}{c}
\text{conditional local}\\[-1mm]
\text{requirements}
\end{array}}.
\]
This scheme captures the additional capability introduced at level
$\Pi C_2$; the detailed activation-aware composition procedure and its
algorithmic realisation are developed in Section~\ref{sec:activation-aware}.

Exclusive choice is the principal motivating case, but $\Pi C_2$ is not
defined by the presence of an XOR operator. Its defining condition is
more general: whenever structural inclusion of a behavioural component
does not imply unconditional execution, Pattern Composition must separate
the local semantics of that component from the activation semantics
imposed by its context.

\subsection{$\Pi C_3$: Iteration-Aware Pattern Composition}

Iteration introduces a further qualitative change in Pattern
Composition. Activation-aware composition distinguishes between
components that are structurally present and those that are required to
execute. Iteration introduces a different problem: the same behavioural
component may be executed repeatedly, and some specification tasks may
require these executions to be distinguished from one another.

Not every iterative structure requires such a distinction. Many
properties of repetitive behaviour can be expressed without identifying
individual executions. Temporal properties concerning recurrence,
response, progress, or repeated reachability may treat successive
executions uniformly. This view is consistent with the classical
temporal treatment of reactive and repeatedly evolving
behaviour~\cite{Manna-Pnueli-1992}. Explicit occurrence identity becomes
necessary only when the intended analysis must distinguish individual
executions or relate information belonging to different executions.

For this reason, $\Pi C_3$ does not automatically introduce
occurrence-sensitive reasoning for every loop encountered in a pattern
expression. Instead, the relevant iterative structure is selected
externally. Let
\[
Iter(E)
\]
denote the set of iterative pattern instances occurring in a pattern
expression $E$. The user supplies a selection
\[
\rho \subseteq Iter(E)
\]
identifying those iterative instances for which individual executions
are relevant. This distinction is important when several loops occur in
the same expression, or when the same pattern occurs in more than one
iterative position: the selection refers to a particular structural
instance rather than merely to a pattern type.

The distinction between the preceding level and $\Pi C_3$ can therefore
be expressed as
\[
\begin{array}{rcl}
\Pi C_2 &:&
\textit{Under what conditions is a pattern required to execute?}\\[1mm]
\Pi C_3 &:&
\textit{How are individual executions of a selected iterative pattern
distinguished and related?}
\end{array}
\]
The latter question arises only when occurrence-sensitive reasoning has
been explicitly requested.

Let $\nu\in\rho$ denote a selected iterative pattern instance. The new
information introduced by $\Pi C_3$ is an \emph{occurrence identity}.
We denote the $i$-th execution associated with $\nu$ schematically by
\[
Occ(\nu,i).
\]
The two arguments have different roles: $\nu$ identifies the structural
instance selected in the pattern expression, whereas $i$ distinguishes
one execution of that instance from another. Thus,
\[
\boxed{\text{structural instance identity}}
\qquad\text{and}\qquad
\boxed{\text{execution occurrence identity}}
\]
remain explicitly separated.

Occurrence identity can subsequently qualify the behavioural interface
of the selected pattern. If $P_\nu$ denotes the pattern occurring at
the selected iterative position $\nu$, its otherwise undifferentiated
interface
\[
P_\nu.ini,
\qquad
P_\nu.fin
\]
can be refined, schematically, to
\[
P_\nu.ini(i),
\qquad
P_\nu.fin(i),
\]
under the occurrence condition $Occ(\nu,i)$. The occurrence identifier
does not introduce a new behavioural pattern; it identifies a particular
execution of the same pattern and allows the generated specification to
preserve this distinction.

This qualification makes two forms of reasoning possible. Properties
may relate events belonging to the same occurrence, for example
\[
P_\nu.ini(i)
\quad\leadsto\quad
P_\nu.fin(i),
\]
or they may relate different occurrences of the selected iterative
structure, for example
\[
Occ(\nu,i)
\quad\prec\quad
Occ(\nu,i+1).
\]
The notation is deliberately schematic at this level. Its purpose is to
identify the additional compositional information required by
occurrence-sensitive reasoning; the precise logical encoding is
developed separately.

Importantly, iteration itself does not imply a change of logical
formalism. If individual executions need not be distinguished, the
iterative behaviour may remain within the temporal specification
mechanisms used at the preceding levels. Even a fixed and bounded set of
distinguished executions can, in principle, be represented by explicit
occurrence qualification. A different representational problem arises
when reasoning must range symbolically over an unbounded family of
individually identifiable occurrences. In that case, explicit occurrence
identifiers and relations between them may motivate a richer logical
representation, potentially involving first-order or first-order
temporal machinery. The need for such an extension follows from the
required occurrence-sensitive reasoning, not from iteration alone.

At the algorithmic level, $\Pi C_3$ extends the preceding composition
mechanisms by adding occurrence qualification only to user-selected
iterative structures. Its essential processing scheme is
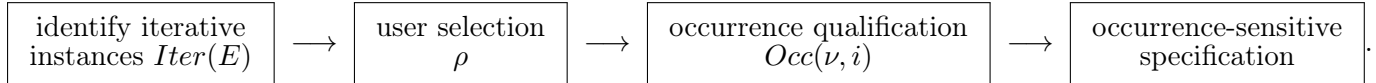
\[
\boxed{\begin{array}{c}
\text{identify iterative}\\[-1mm]
\text{instances }Iter(E)
\end{array}}
\;\longrightarrow\;
\boxed{\begin{array}{c}
\text{user selection}\\[-1mm]
\rho
\end{array}}
\;\longrightarrow\;
\boxed{\begin{array}{c}
\text{occurrence qualification}\\[-1mm]
Occ(\nu,i)
\end{array}}
\;\longrightarrow\;
\boxed{\begin{array}{c}
\text{occurrence-sensitive}\\[-1mm]
\text{specification}
\end{array}}.
\]

Accordingly, $\Pi C_3$ can be characterised schematically as
\[
\Pi C_3 :
(\mathcal{E}_3,\rho)
\longrightarrow
2^{\mathcal{L}},
\]
where $\mathcal{E}_3$ admits iterative behavioural structures and
$\rho\subseteq Iter(E)$ identifies those structural instances for which
occurrence-sensitive treatment is required. For iterative components not
selected by $\rho$, explicit occurrence identity is not introduced.

The transition from $\Pi C_2$ to $\Pi C_3$ therefore does not mean that
every loop is unfolded into separately identified executions. Rather,
$\Pi C_3$ adds a new compositional capability---\emph{occurrence
qualification}---that can be applied selectively when the intended
analysis requires individual executions of an iterative structure to be
distinguished. Occurrence identity is introduced on demand, not by
default.

\subsection{$\Pi C_4$: Explainable Pattern Composition}

The preceding levels establish how patterns are instantiated ($\Pi C_0$), 
abstracted ($\Pi C_1$), activated ($\Pi C_2$), and distinguished 
($\Pi C_3$). None of these levels, however, preserves information about 
why a particular formula was generated. This becomes problematic when 
the generated specification is used for automated reasoning: a prover 
may return an unsatisfiable core or a set of involved formulae, but 
such results are expressed at the logical level and no longer directly 
reveal which elements of the original behavioural structure produced 
them. This challenge has been identified in recent work on explainable 
verification of hierarchical workflows: ``it remains unclear why such 
models satisfy or violate logical properties, or how individual 
elements contribute to overall behavior''~\cite{Klimek-Blazowski-2026-ASE}.

This problem mirrors challenges in process mining, where discovered 
models are often too complex to interpret (the ``spaghetti model'' 
problem)~\cite{Peeva-etal-2022}, discovery algorithms operate 
as black boxes that do not explain their 
results~\cite{Piccirilli-etal-2025-PINPOINT,Hoogendoorn-etal-2023}, 
and analysts struggle to relate model elements to 
their organisational context. Explainable Process 
Mining (XPM) has emerged to address these concerns by integrating 
XAI principles into process mining techniques~\cite{Mukala-2026}. 
Similarly, analytic provenance has been proposed to support validation, 
reproducibility, and reuse of analysis results in exploratory process 
mining~\cite{Zerbato-etal-2023}. $\Pi C_4$ generalises these insights to 
the level of logical specification.

\subsubsection{Provenance Data Structure}

For each generated formula $\varphi$, the provenance record is a tuple:
\[
Prov(\varphi) = \langle sid, rules, ctx, deps \rangle,
\]
where:
\begin{itemize}
    \item $sid \in \mathcal{S}$ is a unique structural identifier
          assigned to the pattern instance that generated $\varphi$
          (e.g., $P_{1.2.1}$ for the third child of the second child
          of the root);
    \item $rules \subseteq \mathcal{R}$ is the set of composition
          rules applied to derive $\varphi$ (e.g., 
          $\{SEQ\_PROP, XOR\_ACT, LOOP\_OCC\}$);
    \item $ctx \in \mathcal{C}$ captures the context at generation
          time (e.g., iteration index $i$, branch $b \in \{1,2\}$,
          nesting depth $d$);
    \item $deps \subseteq \mathcal{F}$ is the set of formulae on which
          $\varphi$ depends (e.g., $\{Act(P,i), Act(Q,i)\}$).
\end{itemize}

\subsubsection{Instrumentation of Composition Rules}

To populate provenance records, each composition rule is instrumented.
The instrumentation is defined uniformly for all rules and takes the
form of a wrapper that intercepts formula generation. For a rule $r$
with input patterns $P_1,\ldots,P_k$ and output formula $\varphi$,
the instrumented rule:

\begin{enumerate}
    \item Computes $sid$ as the structural identifier of the pattern
          instance being processed;
    \item Computes $ctx$ from the current execution context (iteration
          index, branch identifier, etc.);
    \item Records $r$ in $rules$;
    \item Sets $deps$ to the set of formulae generated by the input
          patterns;
    \item Returns $\varphi$ annotated with the resulting provenance
          record.
\end{enumerate}

This instrumentation is modular: rules are wrapped without modifying
their core logic, and provenance is accumulated as composition proceeds.

\subsubsection{Provenance Propagation Rules}

The following table defines how provenance is propagated through the
composition hierarchy. Each rule specifies the provenance transformation
that must be applied when combining sub-expressions.

\[
\begin{array}{|c|c|}
\hline
\text{Composition Rule} & \text{Provenance Transformation} \\
\hline
SEQ(P,Q) & 
\begin{array}{l}
Prov(SEQ) = \langle P.sid \cup Q.sid, \{SEQ\_PROP\}, ctx, \\
\quad \{Prov(P), Prov(Q)\} \rangle
\end{array} \\
\hline
XOR(P,Q) & 
\begin{array}{l}
Prov(XOR) = \langle P.sid \cup Q.sid, \{XOR\_ACT\}, ctx, \\
\quad \{Prov(P), Prov(Q)\} \rangle
\end{array} \\
\hline
LOOP(P) & 
Prov(LOOP) = \langle P.sid, \{LOOP\_OCC\}, \{i\}, \{Prov(P,i)\} \rangle \\
\hline
CONS\_EX(P) & 
Prov(CONS\_EX(P)) = \langle P.sid, \{CONS\_EX\_ABS\}, ctx, \emptyset \rangle \\
\hline
\end{array}
\]

For example, for an expression $Loop(XOR(Seq(P,Q), R))$, the provenance
of the formula $Act(P,i)$ is:

\[
Prov(Act(P,i)) = \langle P_{1.1.1}, 
   \{SEQ\_PROP, XOR\_ACT, LOOP\_OCC\}, 
   \{i\}, 
   \emptyset \rangle.
\]

This trace reveals that $P$ appears in the first branch of XOR
(position $1.1.1$), which is nested inside a loop, and that the formula
was derived through three composition rules.

\subsubsection{Provenance Consolidation and Query Interface}

For human interpretation, provenance records are consolidated into
higher-level explanations. The consolidation process applies the
following transformations:
\begin{enumerate}
    \item \textbf{Grouping by source}: All formulae originating from
          the same $sid$ are grouped together.
    \item \textbf{Path compression}: For each formula, the sequence of
          rules in $rules$ is summarised as a path in the composition
          tree (e.g., "root $\to$ XOR $\to$ Seq $\to$ P").
    \item \textbf{Contextual annotation}: For each formula, the context
          $ctx$ is translated into a human-readable description
          (e.g., "inside loop iteration 3, branch 1").
\end{enumerate}

The provenance system supports the following queries, formalised as
functions over the provenance space:

\[
\begin{array}{rcl}
\texttt{source}(\varphi) & = & \{\, sid \mid \langle sid, \_, \_, \_ \rangle = Prov(\varphi)\,\} \\
\texttt{rules}(\varphi) & = & \{\, r \mid \langle \_, rules, \_, \_ \rangle = Prov(\varphi) \,\} \\
\texttt{context}(\varphi) & = & \{\, c \mid \langle \_, \_, ctx, \_ \rangle = Prov(\varphi) \,\} \\
\texttt{deps}(\varphi) & = & \{\, \psi \mid \langle \_, \_, \_, deps \rangle = Prov(\varphi), \psi \in deps \,\} \\
\texttt{trace}(\varphi) & = & \texttt{rules}(\varphi) \cup \texttt{context}(\varphi) \cup \{\texttt{source}(\varphi)\} \\
\end{array}
\]

\subsubsection{Integration with Automated Reasoning}

When a theorem prover returns a set of formulae $C$ (e.g., an
unsatisfiable core), provenance enables projection back to the
behavioural structure:
\[
\texttt{project}(C) = \bigcup_{\varphi \in C} \texttt{trace}(\varphi).
\]

This projection returns the set of structural identifiers, composition
rules, and contexts that contributed to the detected issue. For
example, if an inconsistency is detected, the analyst can immediately
identify which pattern instances are involved and which composition
decisions led to the conflict. This directly addresses the challenge 
identified in the literature on explainable verification of hierarchical 
workflows~\cite{Klimek-Blazowski-2026-ASE}.

\subsubsection{Illustrative Example}

The central question of $\Pi C_4$ is therefore:
\[
\boxed{
\textit{How can reasoning results over the generated specification
be traced back to the behavioural structures that produced them?}
}
\]
This question directly addresses the challenge identified in recent work 
on explainable verification of hierarchical workflows: ``it remains 
unclear why such models satisfy or violate logical properties, or how 
individual elements contribute to overall behavior''~\cite{Klimek-Blazowski-2026-ASE}.

By introducing provenance as a first-class artefact, $\Pi C_4$ enables:
\begin{itemize}
    \item \emph{Traceability}: every formula can be traced back to
          its structural origins via $sid$;
    \item \emph{Diagnosability}: verification failures can be
          explained through $\texttt{trace}$ and $\texttt{project}$;
    \item \emph{Trust}: analysts can query provenance to understand
          and validate the composition process;
    \item \emph{Feedback for AI}: provenance provides structured
          feedback for $\Pi C_{Assist}$, enabling agents to refine
          their proposals based on specific structural issues.
\end{itemize}

$\Pi C_4$ therefore extends Pattern Composition from specification
generation towards explainable specification generation.

\subsection{Summary of the Composition Hierarchy}

The hierarchy introduced above organises Pattern Composition according
to the additional information that must be preserved or introduced as
the behavioural structure becomes more demanding. The levels are
cumulative: each procedure retains the capabilities of its predecessor
and adds a mechanism required by a qualitatively new composition
problem.

The resulting progression can be summarised as follows:
\[
\begin{array}{c|l|l}
\textbf{Level}
&
\textbf{Composition problem}
&
\textbf{Additional mechanism}
\\
\hline
\Pi C_0
&
\text{flat pattern instantiation}
&
\text{direct instantiation}
\\
\Pi C_1
&
\text{hierarchical nesting}
&
\text{structural abstraction }(ConsEx)
\\
\Pi C_2
&
\text{conditional execution}
&
\text{activation qualification }(Act)
\\
\Pi C_3
&
\text{repeated execution}
&
\text{occurrence qualification }(Occ)
\\
\Pi C_4
&
\text{explainability of composition}
&
\text{provenance preservation }(Prov)
\end{array}
\]

Thus, the hierarchy can be read as a sequence of increasingly rich
composition capabilities:
\[
\boxed{\text{instantiate}}
\;\longrightarrow\;
\boxed{\text{abstract}}
\;\longrightarrow\;
\boxed{\text{qualify activation}}
\;\longrightarrow\;
\boxed{\text{qualify occurrence}}
\;\longrightarrow\;
\boxed{\text{preserve provenance}}.
\]

The progression is not intended merely as a classification of workflow
constructs. Each transition identifies information that is absent at
the preceding level but becomes necessary for preserving the intended
behavioural semantics during composition. In this sense, the hierarchy
characterises increasing demands placed on the composition procedure,
rather than increasing syntactic complexity alone.

The $\Pi C$ hierarchy remains orthogonal to the hierarchy of temporal
fragments introduced earlier. A composition procedure determines
\emph{how} behavioural structures are transformed compositionally,
whereas $TL_j$ determines the temporal expressive resources available
for representing the resulting properties. A particular configuration
can therefore still be characterised by
\[
\Pi C_i[TL_j],
\]
without assuming that increasing composition complexity necessarily
requires increasing temporal expressiveness.

This separation is important because the two dimensions need not evolve
together. A structurally demanding composition problem may still
generate properties within a comparatively simple temporal fragment,
whereas a structurally simple pattern may require a temporally richer
property. The framework therefore treats composition capability and
logical expressiveness as independent dimensions whose interaction can
be investigated explicitly.

Finally, the hierarchy is intentionally open-ended. The levels
$\Pi C_0$ through $\Pi C_4$ capture the compositional requirements
identified in the present framework; further levels may be introduced
when additional classes of behavioural structures require qualitatively
new compositional information.

\subsection{An Orthogonal Extension: $\Pi C_{Assist}$}

The hierarchy $\Pi C_0$--$\Pi C_4$ assumes that pattern expressions 
are given explicitly by the analyst. In practice, however, behavioural 
requirements are often expressed in natural language, and identifying 
the appropriate pattern structure requires interpretation. Recent 
advances in large language models (LLMs) and generative AI have 
enabled the automatic generation of behavioural artefacts from 
natural-language descriptions, offering a promising avenue for 
reducing the effort required to produce initial modelling 
results~\cite{Wael-etal-2025-flowfsm}.

The extension introduced in this section is not a successor to 
$\Pi C_4$, but an orthogonal mode of interaction with the hierarchy. 
Its purpose is not to introduce additional compositional information 
beyond that captured by $\Pi C_0$--$\Pi C_4$, but to enable AI 
systems to propose, refine, and validate pattern expressions under 
the guidance of the composition hierarchy.

\subsubsection{Interaction Model}

$\Pi C_{Assist}$ operates through a feedback loop between 
an AI agent and the composition hierarchy:

\begin{enumerate}
    \item \textbf{Proposal}: The AI agent receives a natural-language 
          description of a behavioural requirement and proposes a 
          candidate pattern expression (e.g., $XOR(Seq(P,Q), R)$).
    \item \textbf{Verification}: The candidate is verified against 
          $\Pi C_0$--$\Pi C_4$ conditions: structural well-formedness 
          ($\Pi C_0$), nested abstraction ($\Pi C_1$), activation 
          consistency ($\Pi C_2$), occurrence soundness ($\Pi C_3$), 
          and provenance traceability ($\Pi C_4$).
    \item \textbf{Feedback}: If verification fails, the hierarchy 
          returns structured feedback indicating which condition was 
          violated and which structural elements are responsible.
    \item \textbf{Refinement}: The AI agent revises its proposal based 
          on the feedback and repeats the loop.
\end{enumerate}

\begin{figure}[htb]
\centering
\begin{tikzpicture}[
    node distance=1.2cm and 1.8cm,
    box/.style={rectangle, draw, thick, rounded corners=2mm, 
                minimum width=2.8cm, minimum height=0.9cm, 
                align=center, font=\small},
    arrow/.style={->, >=stealth, thick, shorten >=2pt, shorten <=2pt},
    label/.style={font=\footnotesize\itshape}
]

\node[box, fill=blue!10] (input) {Natural-Language\\Requirements\\or Event Logs};
\node[box, fill=green!10, right=of input] (propose) {Propose\\Pattern Expression};
\node[box, fill=orange!10, right=of propose] (verify) {Verify\\$\Pi C_0$--$\Pi C_4$};
\node[box, fill=red!10, below=of verify] (feedback) {Structured\\Feedback};
\node[box, fill=green!10, below=of propose] (refine) {Refine\\Expression};

\draw[arrow] (input) -- node[above, label] {Extract} (propose);
\draw[arrow] (propose) -- node[above, label] {Check} (verify);

\draw[arrow] (verify) -- node[right, label] {Fail} (feedback);
\draw[arrow] (feedback) -- node[below, label] {Guide} (refine);
\draw[arrow] (refine) -- node[right, label] {Update} (propose);

\draw[arrow, dashed, thick] (verify.east) -- ++(1.8,0) 
    node[right, text width=2.5cm, font=\small\itshape, align=center] {Verified\\Specification};

\end{tikzpicture}
\caption{Interaction model for $\Pi C_{Assist}$. The AI agent proposes a pattern expression based on natural-language requirements or event logs, the hierarchy verifies it, and structured feedback guides iterative refinement.}
\label{fig:assist-loop}
\end{figure}
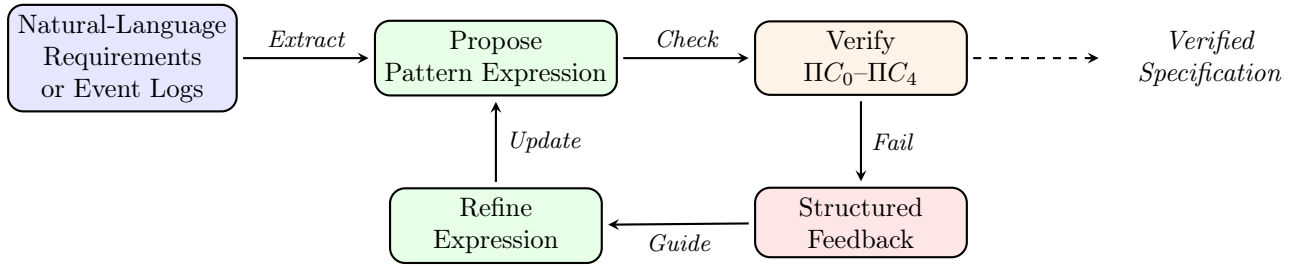

This interaction model is illustrated in Figure~\ref{fig:assist-loop}.

\subsubsection{Relationship to the Hierarchy}

$\Pi C_{Assist}$ is neither a successor to $\Pi C_4$ nor a 
replacement for any of the existing levels. It is an orthogonal 
extension that operates \emph{over} the hierarchy rather than 
\emph{within} it. The hierarchy provides the verification machinery; 
$\Pi C_{Assist}$ provides the AI-assisted interface to that 
machinery.

The central question of this level is therefore:

\[
\boxed{
  \parbox{0.9\columnwidth}{\centering\textit{
    How can AI systems propose, refine, and validate pattern
    expressions under the~guidance of the composition hierarchy?
  }}
}
\]

\subsubsection{Relationship to Explainability}

The provenance mechanism introduced in $\Pi C_4$ is essential for 
$\Pi C_{Assist}$. When a candidate expression fails 
verification, the AI agent needs to know \emph{which} structural 
element caused the failure and \emph{why}. Provenance provides this 
information in a structured form that can be consumed by the agent, 
enabling targeted refinement rather than blind resubmission.

This distinguishes $\Pi C_{Assist}$ from purely conversational 
AI-assisted modelling, where feedback is typically provided in natural 
language and may lack the precision required for systematic correction.

\subsubsection{Summary}

$\Pi C_{Assist}$ extends the applicability of Pattern 
Composition to settings where pattern expressions are not given 
explicitly but must be derived from natural-language requirements. 
It introduces no new compositional information, but enables AI 
systems to interact with the hierarchy through a feedback loop 
mediated by provenance.

\section{Activation-Aware Pattern Composition}
\label{sec:activation-aware}

The transition from $\Pi C_1$ to $\Pi C_2$ addresses a limitation that
cannot be resolved by structural abstraction alone. A nested pattern may
be structurally present in a pattern expression while its execution
depends on the behavioural context imposed by its enclosing pattern.
Pattern composition must therefore preserve two different kinds of
information: the representation of a nested pattern at the composition
boundary and the condition under which that pattern participates in a
particular execution.

The central distinction is
\[
\boxed{
  \text{structural presence}
  \not\equiv
  \text{behavioural activation}
}.
\]
Structural presence is determined by the pattern expression and handled
by hierarchical composition through consolidated expressions. Behavioural
activation, in contrast, is determined by the execution semantics of the
enclosing pattern.

Consider, for example,
\[
XOR(P,Q).
\]
Both $P$ and $Q$ are structurally present and must therefore be represented
during composition. Their structural presence does not, however, imply
that both are executed in the same run: the semantics of $XOR$ activates
exactly one of its alternatives. The same issue arises when only one
alternative is a nested pattern, as in $XOR(P,a)$: the presence of $P$ in
the expression must not by itself require the execution of $P$.

The principle underlying $\Pi C_2$ is consequently
\[
\boxed{
\text{a nested pattern specifies how it behaves;}
\quad
\text{the enclosing pattern determines whether it is activated}
}.
\]
Activation-aware Pattern Composition therefore extends $\Pi C_1$ rather
than replacing it. Consolidated expressions continue to provide the
structural representation of nested patterns, while activation introduces
the execution context in which their execution-dependent requirements
become applicable.

\subsection{Activation Semantics}

Let $\nu$ denote a particular pattern occurrence in an expression, and
let $\alpha_\nu$ denote the condition under which this occurrence
participates in the execution under consideration. Activation must be
distinguished from structural membership. For a pattern expression $E$,
\[
\nu \in E
\]
states that the instance occurs structurally in $E$, whereas
\[
\alpha_\nu
\]
states that it is activated in a particular execution. Activation is
therefore contextual rather than intrinsic to a pattern: the same pattern
specification may occur in different enclosing structures without
changing its internal behavioural meaning, while the condition under
which its execution is required may differ.

To make the relationship between activation and the pattern interface
explicit, define an activation representative for a compositional
argument $x$:
\[
\rho(x) =
\begin{cases}
  \mathrm{Exist}(x), & \text{if $x$ is an atomic activity},\\
  \mathrm{Exist}(Q_x.ini), & \text{if $x$ is an instance of pattern $Q_x$}.
\end{cases}
\]
Thus, for a nested pattern occurrence $\nu$ with schema $P_\nu$, its
interface-level activation representative is
$\rho(\nu)=\mathrm{Exist}(P_\nu.ini)$. The representative connects the
parent's activation semantics to the event-level interface of its
arguments. The root occurrence $r$ is active by convention,
$\alpha_r \equiv \top$; its initial occurrence is required once for the
complete expression, as specified in the recursive procedure below.

The pattern schema can expose the activation semantics of its arguments
as part of its interface:
\[
P = \langle op, Args, Spec, ini, fin, ActRel \rangle,
\]
where $op$ denotes the pattern operator, $Args$ its formal arguments,
$Spec$ its local temporal specification, and $ini$ and $fin$ its initial
and final interface conditions. $ActRel$ specifies how the activation of
the pattern constrains the activation representatives of its arguments.
This makes the activation requirements available to the composition
procedure without changing the local behavioural meaning of the pattern.

Consider, for example,
\[
Seq(P,Q)
\]
and
\[
XOR(P,Q).
\]
If the enclosing sequence is activated, both components are required to
participate in its execution. For a sequential pattern with arguments
$x_1,\ldots,x_k$, the corresponding relation is
\[
ActRel_{Seq}(\alpha,\rho(x_1),\ldots,\rho(x_k))
=
\{\rho(x_i)\leftrightarrow\alpha \mid 1\leq i\leq k\}.
\]
For an exclusive choice, an active parent selects exactly one argument,
whereas an inactive parent selects none:
\[
\begin{aligned}
ActRel_{XOR}(\alpha,\rho(x_1),\ldots,\rho(x_k))
={}&
\{\alpha\rightarrow
  ExactlyOne(\rho(x_1),\ldots,\rho(x_k))\}\\
&{}\cup
\{\rho(x_i)\rightarrow\alpha \mid 1\leq i\leq k\}.
\end{aligned}
\]
For the expression $XOR(P,c)$, with the choice as the root, this relation
requires
\[
ExactlyOne(\mathrm{Exist}(P.ini),\mathrm{Exist}(c)).
\]
If $P$ is selected, its own activation relation governs its arguments;
if $c$ is selected, $P.ini$ is not required. More generally, for a
parent pattern $P_\nu$ with arguments $x_1,\ldots,x_k$, let
\[
A_\nu =
ActRel_{P_\nu}
  \bigl(\alpha_\nu,\rho(x_1),\ldots,\rho(x_k)\bigr)
\]
denote the activation constraints induced by its semantics. Mandatory
structures require all relevant arguments to participate, whereas
alternative structures select among them.

This separation captures the central principle of activation-aware
composition:
\[
\boxed{
\text{a pattern specifies what holds when it executes;}
\quad
\text{its context determines whether it is required to execute}
}.
\]
Structural abstraction is constructed bottom-up:
\[
\boxed{
\text{child}
\longrightarrow
ConsEx(\text{child})
\longrightarrow
\text{parent}
},
\]
whereas activation context originates in the enclosing structure and
flows towards its components:
\[
\boxed{
\text{parent}
\longrightarrow
\text{activation of its arguments}
}.
\]
The two directions therefore carry different compositional information:
\[
\begin{array}{rcl}
\text{bottom-up} &:&
\text{structural abstraction and } ConsEx,\\[1mm]
\text{parent-to-child} &:&
\text{execution context and activation}.
\end{array}
\]
Hierarchical abstraction tells the parent how a nested pattern is
represented compositionally; activation semantics determines under which
execution context that pattern's execution-dependent requirements become
applicable. These different directions do not require separate
composition algorithms: they can be handled within one recursive
procedure.

\subsection{Activation-Aware Logical Specifications}

The need for activation-aware composition becomes visible when a local
pattern specification contains a requirement that can itself force the
pattern to occur. Suppose that the local specification of $P$ contains
the existence requirement
\[
\mathrm{Exist}(P.ini).
\]
If $P$ occurs in a mandatory context, requiring this occurrence is
consistent with the semantics of the enclosing structure. If $P$ occurs
as an alternative branch, however, the same requirement must not force
$P$ to execute independently of whether that branch has been selected.

For example, if both arguments of
\[
XOR(P,Q)
\]
are nested patterns and their local specifications contribute
unconditionally
\[
\mathrm{Exist}(P.ini)
\qquad\text{and}\qquad
\mathrm{Exist}(Q.ini),
\]
the resulting specification may require both alternatives to occur,
while the enclosing pattern permits exactly one of them to be activated.
Even when only one alternative is nested, as in $XOR(P,a)$, an
unconditional requirement $\mathrm{Exist}(P.ini)$ may implicitly force
the selection of $P$.

The problem is therefore not exclusive choice itself, but the presence
of local constraints that impose execution independently of the
activation context of the nested pattern. We call such constraints
\emph{activation-producing}. Other constraints merely describe behaviour
once the pattern is executing and do not themselves require its
execution. We call these \emph{activation-neutral} constraints.

Activation-producing constraints must be conditional on the activation
of the corresponding pattern. In an explicit-activation presentation,
this qualification takes the form
\[
\mathrm{Exist}(P.ini)
\qquad\leadsto\qquad
\alpha_P\rightarrow \mathrm{Exist}(P.ini).
\]
In the interface-based presentation used here, the standalone initial
occurrence of a nested pattern is instead represented by its activation
representative,
$\rho(P)=\mathrm{Exist}(P.ini)$, and is constrained by the enclosing
pattern's $ActRel$. It is therefore not added a second time as an
unconditional local requirement. Activation-producing constraints
other than this standalone initial-occurrence requirement remain subject
to qualification.

This qualification should not be applied indiscriminately to the whole
local specification. Consider, for example,
\[
\Box(P.ini\rightarrow\Diamond P.fin).
\]
This formula constrains the behaviour of $P$ once its initial condition
occurs, but it does not itself require $P.ini$ to occur. If $P$ is not
activated and $P.ini$ never holds, the formula does not force execution.
The same observation applies to a safety property such as
\[
\Box\neg(P.ini\land P.fin),
\]
which restricts possible behaviour without requiring either interface
condition to occur.

Let
\[
AP(P)\subseteq P.pat()
\]
denote the set of activation-producing formulae of a pattern instance
$P$, excluding the standalone initial-occurrence requirement when that
requirement is represented by $\rho(P)$. For a local specification $L$
and an activation condition $\alpha$, define
\[
Qual(L,\alpha)
=
\bigl(L\setminus AP(L)\bigr)
\cup
\left\{
  \alpha\rightarrow\varphi
  \;\middle|\;
  \varphi\in AP(L)
\right\},
\]
where $AP(L)$ denotes the activation-producing constraints contained in
$L$. The activation-aware local specification is then
\[
P.pat^{Act}()
=
Qual(P.pat(),\alpha_P).
\]
For a legacy pattern specification that includes
$\mathrm{Exist}(P.ini)$ as a local formula, this formula is activation-
producing and must be qualified as above. In the interface-based
formulation, the same occurrence condition is supplied through
$\rho(P)$ and the parent relation, and must not be imposed independently
a second time.

This formulation makes $\Pi C_2$ deliberately conservative. Formulae
that do not create an execution obligation remain unchanged; activation
qualification is introduced only where a local specification would
otherwise force execution independently of context. Well-composedness,
consolidated expressions, and activation therefore retain distinct
roles: well-composedness concerns internal behavioural coherence,
$ConsEx$ provides structural abstraction, and activation controls the
applicability of execution-producing requirements.

\subsection{The $\Pi C_2$ Algorithm}

The $\Pi C_2$ procedure extends hierarchical Pattern Composition by
propagating activation context through the pattern-expression tree,
using the interface-level activation representatives of compositional
arguments, and qualifying any additional local constraints that may
force execution. Let $E$ be a hierarchical pattern expression
represented by a rooted tree. For a pattern instance $\nu$, let
\[
P_\nu
\]
denote its pattern schema, let
\[
args(\nu)=(x_1,\ldots,x_k)
\]
denote its ordered compositional arguments, and let $nested(\nu)$ be
the subset consisting of pattern instances. For a nested child
$\mu_i$, its activation condition is represented by
\[
\alpha_{\mu_i}\equiv\rho(\mu_i)
=\mathrm{Exist}(P_{\mu_i}.ini).
\]
For an atomic argument $x_j$, its representative is
$\rho(x_j)=\mathrm{Exist}(x_j)$.

For a local specification $L$ and activation condition $\alpha$, define
the qualification operator
\[
Qual(L,\alpha)
=
\bigl(L\setminus AP(L)\bigr)
\cup
\left\{
  \alpha\rightarrow\varphi
  \;\middle|\;
  \varphi\in AP(L)
\right\},
\]
where $AP(L)$ contains the activation-producing constraints in $L$,
excluding a nested pattern's standalone initial-occurrence requirement
when it is already represented by $\rho$. Activation relations between
a parent and its arguments are determined by the semantics of the parent
pattern:
\[
A_\nu =
ActRel_{P_\nu}
\bigl(\alpha_\nu,\rho(x_1),\ldots,\rho(x_k)\bigr).
\]

The recursive procedure returns a specification $S_\nu$ and a
consolidated expression $C_\nu$ for each subtree:
\[
\boxed{
\Pi C_2(\nu,\alpha_\nu)
\longrightarrow
(S_\nu,C_\nu)
},
\]
where $S_\nu$ is the specification generated for the subtree rooted at
$\nu$, and $C_\nu$ is its consolidated expression.

\begin{enumerate}
  \item Let
  \[
  args(\nu)=(x_1,\ldots,x_k)
  \]
  be the ordered arguments of $P_\nu$, and let $nested(\nu)$ be the
  nested pattern instances among them.

  \item For every argument $x_j$, use its activation representative
  $\rho(x_j)$. Construct the activation constraints
  \[
  A_\nu =
  ActRel_{P_\nu}
  \bigl(\alpha_\nu,\rho(x_1),\ldots,\rho(x_k)\bigr).
  \]
  For sequential composition, these constraints require every argument
  to participate whenever the parent is active. For exclusive choice,
  they require exactly one argument when the parent is active and no
  argument when it is inactive.

  \item For every nested child $\mu_i\in nested(\nu)$, recursively compute
  \[
  (S_{\mu_i},C_{\mu_i})
  =
  \Pi C_2(\mu_i,\rho(\mu_i)).
  \]

  \item Let $\bar{x}_j$ denote the argument passed to the parent after
  recursive processing:
  \[
  \bar{x}_j =
  \begin{cases}
    C_{\mu_i}, & \text{if $x_j=\mu_i$ for some }
                    \mu_i\in nested(\nu),\\
    x_j, & \text{if $x_j$ is atomic}.
  \end{cases}
  \]
  Instantiate the local specification of $P_\nu$ using these
  consolidated arguments:
  \[
  L_\nu =
  Inst\bigl(P_\nu.pat(),\bar{x}_1,\ldots,\bar{x}_k\bigr).
  \]

  \item Qualify any activation-producing constraints in the local
  specification:
  \[
  L_\nu^{Act}=Qual(L_\nu,\alpha_\nu).
  \]
  The occurrence condition represented by $\rho(\nu)$ is not duplicated
  in $L_\nu$; it is imposed through the parent relation, or once for
  the complete expression when $\nu$ is the root.

  \item Construct the consolidated expression
  \[
  C_\nu =
  ConsEx(P_\nu,\bar{x}_1,\ldots,\bar{x}_k)
  \]
  according to the hierarchical composition rules of $\Pi C_1$.

  \item Add the initial-occurrence requirement only for the root:
  \[
  I_\nu =
  \begin{cases}
    \{\mathrm{Exist}(P_\nu.ini)\}, & \text{if $\nu$ is the root},\\
    \emptyset, & \text{otherwise}.
  \end{cases}
  \]
  Return
  \[
  S_\nu =
  L_\nu^{Act}
  \cup A_\nu
  \cup
  \bigcup_{\mu_i\in nested(\nu)}S_{\mu_i}
  \cup I_\nu,
  \qquad C_\nu.
  \]
\end{enumerate}

For the complete expression $E$ with root $r$, the root is active by
convention:
\[
\alpha_r=\top.
\]
The resulting specification is
\[
\boxed{
\Pi C_2(E)=\pi_1\bigl(\Pi C_2(r,\top)\bigr),
}
\]
where $\pi_1$ selects the specification component of the returned pair.
The root consolidated expression is retained internally for composition
but is not part of the external result.

The procedure combines the two information directions required by
activation-aware composition within a single recursion. Activation
conditions are supplied from parent to child, whereas consolidated
expressions are returned from child to parent:
\[
\boxed{
\alpha_{\mathrm{parent}}
\;\downarrow\;
\rho(\mathrm{child})
\qquad\qquad
ConsEx(\mathrm{child})
\;\uparrow\;
\mathrm{parent}
}.
\]
This directionality explains why hierarchical abstraction and
activation can be handled together without requiring a separate
top-down traversal.

If all nested pattern instances are mandatory, their activation
representatives are equivalent to the activation of their parent.
Since the root is active, every nested activation condition reduces to
$\top$. Consequently,
\[
Qual(L,\top)=L,
\]
and the activation relations reduce to tautologies. The root-only
initial-occurrence requirement is the same as in hierarchical
composition, so
\[
\Pi C_2(E)\equiv\Pi C_1(E).
\]
Thus, $\Pi C_2$ is a conservative extension of hierarchical Pattern
Composition: it adds conditional activation where needed while
preserving the result for mandatory compositions.

\paragraph{Recursive characterization.}
Let $E$ be a rooted pattern-expression tree. For each pattern instance
$\nu$, let $args(\nu)=(x_1,\ldots,x_k)$ be its ordered arguments, and
let $nested(\nu)$ be the subset of arguments that are pattern instances.
The root $r$ is active by convention, $\alpha_r\equiv\top$. For every
nested instance $\nu\neq r$, its activation representative is
$\alpha_\nu\equiv\rho(\nu)$. The procedure returns a specification
$S_\nu$ and a consolidated expression $C_\nu$ for each subtree:
\[
\Pi C_2(\nu,\alpha_\nu)\longrightarrow(S_\nu,C_\nu).
\]
For each argument $x_j$, let $\bar{x}_j=C_{\mu_i}$ if $x_j=\mu_i$ is a
nested instance, and let $\bar{x}_j=x_j$ otherwise. The recursive
construction is defined by
\[
\begin{aligned}
A_\nu
  &\equiv ActRel_{P_\nu}
      \bigl(\alpha_\nu,\rho(x_1),\ldots,\rho(x_k)\bigr),\\
L_\nu
  &\equiv Inst\bigl(P_\nu.pat(),\bar{x}_1,\ldots,\bar{x}_k\bigr),\\
L_\nu^{Act}
  &\equiv Qual(L_\nu,\alpha_\nu),\\
C_\nu
  &\equiv ConsEx(P_\nu,\bar{x}_1,\ldots,\bar{x}_k),\\
I_\nu
  &\equiv
    \begin{cases}
      \{\mathrm{Exist}(P_\nu.ini)\}, & \nu=r,\\
      \emptyset, & \nu\neq r,
    \end{cases}\\
S_\nu
  &\equiv L_\nu^{Act}\cup A_\nu
      \cup\displaystyle\bigcup_{\mu_i\in nested(\nu)}S_{\mu_i}
      \cup I_\nu,
\end{aligned}
\]
where each nested child is processed recursively as
$(S_{\mu_i},C_{\mu_i})=
\Pi C_2(\mu_i,\rho(\mu_i))$. Here, $P_\nu.pat()$ excludes the standalone
initial-occurrence requirement represented by $\rho(\nu)$; any other
activation-producing constraints are qualified by $Qual$. The complete
procedure returns the specification component for the root:
\[
\Pi C_2(E)=\pi_1\bigl(\Pi C_2(r,\top)\bigr).
\]

\section{Evaluation}

\subsection{Research Questions and Experimental Setup}

The evaluation investigates whether activation-aware Pattern Composition
provides the intended semantics for conditionally executed nested
patterns while preserving the behaviour of hierarchical composition in
mandatory contexts. We formulate the following research questions:

\begin{description}

\item[RQ1:] \emph{Does $\Pi C_2$ preserve the intended activation
alternatives in conditionally composed pattern expressions?}

This question evaluates whether activation-aware composition avoids
unintended execution obligations introduced by unconditional local
existence constraints.

\item[RQ2:] \emph{Does $\Pi C_2$ reduce to $\Pi C_1$ when all nested
patterns are mandatory?}

This question evaluates the conservative-extension property of the
proposed procedure.

\item[RQ3:] \emph{What additional specification cost is introduced by
activation-aware composition?}

This question considers the number of activation conditions,
activation-qualified constraints, and generated formulae as the size and
nesting depth of pattern expressions increase.

\end{description}

The evaluation uses systematically constructed families of hierarchical
pattern expressions. The principal cases contain exclusive-choice
structures, since these expose the difference between structural
presence and behavioural activation most directly. Representative
expressions include
\[
XOR(P,Q),
\qquad
XOR(P,a),
\qquad
Seq(XOR(P,Q),R),
\]
\[
XOR(Seq(P,Q),R),
\qquad
XOR(P,XOR(Q,R)),
\]
together with recursively generated variants of increasing nesting
depth.

Mandatory hierarchical expressions based on sequential composition are
used as control cases, for example
\[
Seq(P,Q)
\qquad\text{and}\qquad
Seq(P,Seq(Q,R)).
\]

For each expression $E$, we generate the specification produced by
hierarchical composition, $\Pi C_1(E)$, and the corresponding
activation-aware specification, $\Pi C_2(E)$. Conditional structures are
then examined with respect to three properties.

For an exclusive choice between nested alternatives $P$ and $Q$, we
check:

\[
\neg\bigl(Act(P)\land Act(Q)\bigr),
\]
ensuring exclusiveness, and the satisfiability of
\[
Act(P)\land\neg Act(Q)
\]
and
\[
\neg Act(P)\land Act(Q),
\]
ensuring that both intended alternatives remain feasible.

This distinction is important because satisfiability of the complete
specification alone is insufficient. A specification may remain
satisfiable while an unconditional local requirement such as
$Exist(P.ini)$ eliminates one of the alternatives admitted by the
workflow structure.

For mandatory expressions, we additionally compare the specifications
generated by $\Pi C_1$ and $\Pi C_2$ under
\[
Act(P)=\top
\]
for all relevant nested instances. This provides an empirical check of
the reduction property
\[
\left.\Pi C_2(E)\right|_{Act=\top}
\equiv
\Pi C_1(E).
\]

Finally, for each expression we record its number of pattern instances,
nesting depth, number of generated logical formulae, number of activation
relations, and number of activation-producing constraints qualified by
$\Pi C_2$. These measurements characterise the additional specification
overhead introduced by activation-aware composition.

\subsection{Conditional-Activation Correctness}

Conditional-activation correctness concerns whether the generated
specification preserves the alternatives admitted by the enclosing
workflow structure.

For an exclusive choice
\[
XOR(P,Q),
\]
the intended activation space consists of exactly two admissible cases:
\[
Act(P)\land\neg Act(Q)
\]
and
\[
\neg Act(P)\land Act(Q).
\]
The case
\[
Act(P)\land Act(Q)
\]
must be excluded.

Accordingly, for every exclusive-choice expression considered in the
evaluation, we check the following conditions:

\[
\mathrm{C}_1:
\qquad
\Pi C_2(E)
\models
\neg\bigl(Act(P)\land Act(Q)\bigr),
\]

\[
\mathrm{C}_2:
\qquad
\Pi C_2(E)
\land
Act(P)\land\neg Act(Q)
\quad\text{is satisfiable},
\]

\[
\mathrm{C}_3:
\qquad
\Pi C_2(E)
\land
\neg Act(P)\land Act(Q)
\quad\text{is satisfiable}.
\]

Condition $\mathrm{C}_1$ verifies exclusiveness, whereas
$\mathrm{C}_2$ and $\mathrm{C}_3$ verify that neither admissible branch
has been eliminated by local execution requirements.

The same principle is applied recursively to nested alternatives. If an
exclusive-choice node $\nu$ has children $\mu_1$ and $\mu_2$, then
correct activation requires
\[
Act(\nu)
\rightarrow
\bigl(Act(\mu_1)\oplus Act(\mu_2)\bigr),
\]
while both branch-selecting assignments must remain satisfiable whenever
$Act(\nu)$ holds.

This criterion distinguishes activation correctness from satisfiability
of the global specification alone. Consider, for example,
\[
XOR(P,a),
\]
where the local specification of the nested pattern $P$ contains
\[
Exist(P.ini).
\]
Under ordinary hierarchical composition, the global specification may
remain satisfiable while the unconditional existence requirement forces
the branch containing $P$, thereby eliminating the alternative $a$.
Satisfiability alone would therefore fail to reveal the compositional
problem.

Under $\Pi C_2$, the corresponding requirement becomes
\[
Act(P)\rightarrow Exist(P.ini),
\]
so that the local specification of $P$ constrains its behaviour only
when the branch containing $P$ is activated.

For each evaluated expression, we therefore compare $\Pi C_1$ and
$\Pi C_2$ with respect to:
\[
\boxed{
\text{global satisfiability}
\;+\;
\text{exclusiveness}
\;+\;
\text{feasibility of every intended alternative}
}.
\]

A conditional composition is considered activation-correct only when
all three requirements are satisfied.

\subsection{Conservativity and Specification Growth}

A desirable property of $\Pi C_2$ is that activation awareness should
not modify hierarchical composition when the enclosing structure makes
all nested components mandatory. In such a case, activation introduces
no additional behavioural choice.

Let $E$ be a hierarchical pattern expression in which every nested
pattern instance is mandatory in its enclosing context. Then, for every
pattern instance $\nu$,
\[
\alpha_\nu = \top.
\]
For its local specification $L_\nu$, the qualification operator therefore
satisfies
\[
Qual(L_\nu,\top)
=
(L_\nu\setminus AP_\nu)
\cup
\left\{
\top\rightarrow\varphi
\;\middle|\;
\varphi\in AP_\nu
\right\}
\equiv
L_\nu.
\]

Hence activation qualification does not change the local specification.
After eliminating activation relations that become tautological in a
fully mandatory context, the resulting specification satisfies
\[
\boxed{
\Pi C_2(E)\equiv\Pi C_1(E)
}.
\]

This property expresses conservativity of $\Pi C_2$ with respect to
hierarchical Pattern Composition: activation-aware composition extends
the applicability of $\Pi C_1$ without changing its result for
structures in which conditional activation is absent.

The second evaluation aspect concerns the additional specification
introduced by activation awareness. Let
\[
N(E)
\]
denote the number of pattern instances in $E$, and let
\[
AP(E)
=
\sum_{\nu\in V_E}|AP_\nu|
\]
denote the total number of activation-producing constraints occurring
in their local specifications.

Let
\[
AR(E)
=
\bigcup_{\nu\in V_E} AR_\nu
\]
be the set of activation relations generated for the expression.
Since activation qualification replaces an activation-producing formula
$\varphi$ by
\[
\alpha_\nu\rightarrow\varphi
\]
rather than introducing an additional local formula, the increase in
the number of generated formulae is determined primarily by the
activation relations:
\[
|\Pi C_2(E)|
-
|\Pi C_1(E)|
=
|AR(E)|.
\]

Formula count alone does not capture the complete overhead, because
qualification also increases the syntactic size of activation-producing
constraints. We therefore record separately:

\[
N(E),\qquad
depth(E),\qquad
|AP(E)|,\qquad
|AR(E)|,\qquad
|\Pi C_1(E)|,\qquad
|\Pi C_2(E)|.
\]

For a fixed pattern library with bounded pattern arity and bounded local
specification size, both the number of activation relations and the
number of qualified constraints grow at most linearly with the number
of pattern instances. The evaluation examines whether this expected
behaviour is observed for increasingly nested expression families.

The purpose of this analysis is not primarily to benchmark execution
time, but to determine the logical overhead introduced by conditional
activation and to verify that it remains proportional to the structural
growth of the composed expression.

\subsection{Results}

Table~\ref{tab:activation-results} summarises the results for the
conditional-expression families. For each expression, we report whether
the specification is satisfiable, whether exclusiveness is preserved,
and whether every alternative admitted by the workflow structure remains
feasible.

\begin{table}[htb]
\centering
\caption{Conditional-activation results.}
\label{tab:activation-results}
\begin{tabular}{lcccc}
\hline
Expression &
Method &
SAT &
Exclusive &
All alternatives feasible \\
\hline
$XOR(P,Q)$ &
$\Pi C_1$ &
-- & -- & -- \\

$XOR(P,Q)$ &
$\Pi C_2$ &
-- & -- & -- \\

$XOR(P,a)$ &
$\Pi C_1$ &
-- & -- & -- \\

$XOR(P,a)$ &
$\Pi C_2$ &
-- & -- & -- \\

$Seq(XOR(P,Q),R)$ &
$\Pi C_1$ &
-- & -- & -- \\

$Seq(XOR(P,Q),R)$ &
$\Pi C_2$ &
-- & -- & -- \\

$XOR(P,XOR(Q,R))$ &
$\Pi C_1$ &
-- & -- & -- \\

$XOR(P,XOR(Q,R))$ &
$\Pi C_2$ &
-- & -- & -- \\
\hline
\end{tabular}
\end{table}

\paragraph{RQ1: Conditional-activation correctness.}
The comparison focuses on whether ordinary hierarchical composition can
introduce unintended execution obligations and whether $\Pi C_2$
removes them. In particular, satisfiability is considered together with
branch feasibility: a specification is not regarded as
activation-correct merely because it is satisfiable if one of the
alternatives admitted by the workflow structure has become impossible.

For every exclusive-choice node with alternatives $P$ and $Q$, the
activation-aware specification is checked against
\[
\neg(Act(P)\land Act(Q)),
\]
and against the satisfiability of
\[
Act(P)\land\neg Act(Q)
\]
and
\[
\neg Act(P)\land Act(Q).
\]
The experimental results therefore determine whether $\Pi C_2$
preserves the complete intended activation space rather than only global
consistency.

\paragraph{RQ2: Conservativity.}
For mandatory hierarchical expressions, the specifications generated by
$\Pi C_1$ and $\Pi C_2$ are compared after setting the activation
conditions of all nested instances to $\top$. The expected reduction
property is
\[
\left.\Pi C_2(E)\right|_{Act=\top}
\equiv
\Pi C_1(E).
\]
The result of this comparison determines whether activation-aware
composition changes specifications in cases where no conditional
execution is present.

\paragraph{RQ3: Specification growth.}
Table~\ref{tab:growth-results} reports the structural size of the
evaluated expressions together with the additional logical material
introduced by activation-aware composition.

\begin{table}[t]
\centering
\caption{Specification growth under activation-aware composition.}
\label{tab:growth-results}
\begin{tabular}{lrrrrrr}
\hline
Expression &
$N(E)$ &
depth &
$|AP(E)|$ &
$|AR(E)|$ &
$|\Pi C_1(E)|$ &
$|\Pi C_2(E)|$ \\
\hline
$E_1$ & -- & -- & -- & -- & -- & -- \\
$E_2$ & -- & -- & -- & -- & -- & -- \\
$E_3$ & -- & -- & -- & -- & -- & -- \\
$E_4$ & -- & -- & -- & -- & -- & -- \\
\hline
\end{tabular}
\end{table}

The measurements are used to distinguish two sources of additional
logical structure. Activation-producing constraints are qualified rather
than duplicated, whereas activation relations constitute genuinely new
formulae. Consequently, the principal growth of the specification is
expected to follow the number of activation relations induced by the
pattern-expression structure.

\section{Discussion}

\subsection{Pattern Composition as an Information-Preserving Approach}

The proposed hierarchy treats Pattern Composition as a way of preserving
the information required to derive a global behavioural specification
from reusable local specifications. This perspective differs from
treating composition as the repeated substitution of larger expressions
into a single formula. Substitution alone does not identify which
information about a nested pattern must be exposed to its parent, whether
that pattern participates in a particular execution, or how the generated
formulae can later be related to their sources.

A nested pattern retains
its local specification, while its consolidated expression provides the
abstraction needed at the composition boundary. This separates the
behaviour internal to a pattern from the interface through which the
pattern contributes to its enclosing structure. The resulting
specification can therefore be constructed recursively, from nested
patterns towards the root, without treating a complete nested
specification as an atomic activity.

The correctness of this construction depends on the adequacy of pattern
interfaces and composition rules. A consolidated expression must expose
the information required by its parent, while preserving the meaning of
the nested behaviour. Consequently, the hierarchy does not eliminate the
need to define and validate pattern schemas. It makes their compositional
obligations explicit: each pattern must provide an interface and rules
that are compatible with the structures in which it can occur.

\subsection{Structural Abstraction and Activation Context}

The distinction between structural presence and behavioural activation
illustrates why richer composition requires more than deeper nesting. At
$\Pi C_1$, a nested pattern can be represented compositionally through
$ConsEx$. At $\Pi C_2$, the parent must also determine whether that
pattern participates in the current execution. These mechanisms carry
information in different directions: structural abstraction is assembled
from child to parent, whereas activation is determined by the enclosing
pattern and propagated to its arguments.

This distinction is especially important for alternative structures. In
$XOR(P,Q)$, both patterns are present in the expression, but exactly one
alternative participates in an execution. If the local specifications
unconditionally require the initial occurrences of both patterns, their
composition can contradict the intended choice semantics. The
parent--argument relation $ActRel$ makes the selection explicit, while
activation qualification prevents other local constraints from imposing
execution independently of that selection. Thus, local specifications
continue to describe pattern behaviour, and the enclosing structure
determines when execution-dependent requirements apply.

The interface-based activation representative
$\rho(P)=\mathrm{Exist}(P.ini)$ makes this relationship concrete, but it
also introduces an assumption: the occurrence of $P.ini$ must adequately
represent the participation of $P$ in the enclosing execution. Pattern
libraries must therefore define their initial conditions consistently
with this interpretation. If an initial event can occur without the
corresponding pattern being selected, or if participation is not reflected
by that event, the pattern requires a different activation representative.
Moreover, each control-flow operator needs an appropriate $ActRel$.
The relations for sequential composition and exclusive choice illustrate
the approach; additional operators require their own semantics and
constraints.

\subsection{Compositional Complexity and Temporal Expressiveness}

The separation between $\Pi C_i$ and $TL_j$ distinguishes two sources of
complexity. The composition level characterizes the information needed to
combine workflow structures. The temporal fragment characterizes the
properties that can be expressed in the resulting specification. A
hierarchically complex expression does not necessarily require a more
expressive temporal logic, just as a simple expression may require a
richer fragment to state its desired properties.

The notation $\Pi C_i[TL_j]$ provides a way to discuss these dimensions
jointly, but it does not imply that every pair is automatically
well-defined or useful. A particular composition procedure must generate
formulae supported by the selected fragment, and any translation to
another logic must preserve the relevant semantics. The separation
therefore supports independent analysis of composition and expressiveness
while leaving their compatibility to be established for each
configuration.

\subsection{Scope and Extension of the Hierarchy}

The levels identify different compositional information requirements:
instantiation at $\Pi C_0$, structural abstraction at $\Pi C_1$,
activation context at $\Pi C_2$, occurrence identity at $\Pi C_3$, and
provenance at $\Pi C_4$. The progression is not intended as a measure of
syntactic size. A small expression may require activation information,
whereas a deeply nested expression composed only of mandatory patterns
may not. A new level is justified when a workflow construct requires
information that the preceding levels do not preserve.

The current treatment develops $\Pi C_2$ as a recursive procedure. The
discussion of iteration at $\Pi C_3$ identifies when repeated executions
must be distinguished, rather than requiring every loop to be unfolded
into separate occurrences. Likewise, $\Pi C_4$ identifies provenance as
information needed to connect generated formulae with pattern instances,
composition rules, and execution context. These levels point to further
compositional capabilities, but their full procedures and interactions
remain to be developed. In particular, combining conditional activation,
occurrence-sensitive iteration, and provenance will require rules that
preserve the meaning of each mechanism when they appear together.

\subsection{Evaluation and AI-Assisted Composition}

The evaluation design focuses on properties that can be examined
systematically: whether conditional alternatives remain feasible, whether
$\Pi C_2$ preserves the result of $\Pi C_1$ for mandatory compositions,
and how the generated specification grows with expression size. Constructed
expression families make these checks controlled and repeatable. They do
not, by themselves, establish that the approach scales to realistic
workflow models or that the selected pattern interfaces capture their
intended domain semantics. Evaluation on realistic models is therefore
needed alongside the proposed controlled checks.

The orthogonal mode $\Pi C_{\mathrm{Assist}}$ uses the hierarchy to check
AI-proposed or refined pattern expressions and to return structured
feedback. These checks can identify violations of compositional
conditions, but they do not establish that an expression faithfully
represents its natural-language requirement. That question requires
separate evaluation of the AI-assisted modelling process. The hierarchy
therefore provides a formal basis for checking compositional consistency,
while semantic adequacy remains a distinct concern.


\section{Conclusions}

This paper organizes Pattern Composition as a hierarchy of procedures
that preserve different kinds of compositional information: instantiation,
structural abstraction, activation context, occurrence identity, and
provenance. The hierarchy is independent of the expressive power of the
temporal-logic fragment used to express the resulting specifications.
These dimensions can therefore be studied jointly as $\Pi C_i[TL_j]$.

The main formal contribution is activation-aware Pattern Composition
($\Pi C_2$). Its recursive procedure distinguishes the structural
presence of a nested pattern from its participation in a particular
execution. Parent--argument activation relations express the execution
semantics of enclosing patterns, while consolidated expressions retain
the structural abstractions required for composition. The procedure
preserves hierarchical composition for mandatory structures and supports
conditional execution, with exclusive choice as the principal case.

The hierarchy also identifies occurrence qualification for iterative
patterns ($\Pi C_3$) and provenance for tracing generated formulae
($\Pi C_4$). The paper outlines these capabilities and an orthogonal
AI-assisted mode, $\Pi C_{\mathrm{Assist}}$, in which proposed pattern
expressions are checked against the applicable compositional conditions.
The evaluation protocol defines checks for activation correctness,
conservativity, and specification growth; executing these checks on
larger expression families and realistic workflows remains future work.
Further work will also develop the $\Pi C_3$ procedure, refine the
provenance model into an implemented composition mechanism, and evaluate
whether AI-generated expressions faithfully capture their source
requirements. Compositional validity alone does not establish that
semantic fidelity.

\bibliographystyle{plain}
\bibliography{bib-rk,bib-rk-main,bib-rk-tools}

\end{document}